\documentclass{aa} 

\usepackage{epsf,epsfig,amsmath,amssymb,pifont}
\usepackage{graphicx}
\usepackage{color}
\definecolor{light}{gray}{0.95}
\definecolor{heavy}{gray}{0.35}
    \def\etal{${\rm \hspace*{0.8ex}et\hspace*{0.7ex}al.\hspace*{0.5ex}}$}
    \def\plus{${\rm \hspace*{0.7ex}\&\hspace*{0.7ex}}$}
    \def\ie{i.\,e.\ }
    \def\eg{e.\,g.\ }

\begin{document}

\title{Dust in Brown Dwarfs}

\subtitle{IV. Dust formation and driven turbulence on mesoscopic scales}

\author{Ch. Helling  \inst{1,2,5}
        \and
        R. Klein     \inst{2,3,4}
        \and
        P. Woitke     \inst{5,1} 
        \and
        U. Nowak     \inst{2}
        \and
        E. Sedlmayr  \inst{1}
       }

\offprints{Ch. Helling \protect \\ e-mail: \tt helling@strw.leidenuniv.nl}

\institute{Zentrum f\"ur Astronomie und Astrophysik, Technische
  Universit\"at Berlin, Hardenbergstra{\ss}e~36, D-10623 Berlin \and
  Konrad-Zuse-Zentrum f\"ur Informationstechnik Berlin,
  Takustra{\ss}e~7, D-14195 Berlin \and Fachbereich Mathematik und
  Informatik, Freie Universit\"at Berlin, Arnimallee 2--6, D-14195
  Berlin \and Potsdam Institute for Climate Impact Research,
  Telegrafenberg A31, D-14473 Potsdam,
  \and Sterrewacht Leiden, P.O. Box 9513, 2300 RA Leiden, The Netherlands}

\date{15.10.03 ; 08.04.04 }

\abstract{Dust formation {in brown dwarf atmospheres} is studied
  by utilising a model for driven turbulence in the mesoscopic
  scale regime. We apply a pseudo-spectral method where waves are
  created and superimposed {within} a {limited} wavenumber
  interval. The turbulent kinetic energy distribution follows the
  Kolmogoroff spectrum which is assumed to be the most likely value.
  Such superimposed, stochastic waves may occur in a convectively
  active environment. They cause nucleation fronts and nucleation
  events and thereby initiate the dust formation process which {
  continues until} all condensible material is consumed. Small
  disturbances {are found to} have a large impact on the
  dust forming system. An initially dust-hostile region, which may
  originally be optically thin, becomes optically thick in a patchy
  way showing considerable variations in the dust properties during
  the formation process.  The dust appears in lanes and curls as a
  result of the interaction with waves, \ie turbulence, which form
  larger and larger structures with time. Aiming on a physical
  understanding of the variability of brown dwarfs, related to
  structure formation in substellar atmospheres, we work out first
  necessary criteria for small-scale closure models to be applied
  in macroscopic simulations of dust forming astrophysical systems.}

\maketitle

\sloppy

\section{Introduction}\label{sec:Intro}
Substellar objects like brown dwarfs and (extrasolar) planets are
largely - but not entirely - convective with considerable overshoot in
the upper atmosphere. They also provide excellent conditions for the
gas phase transition to solids or liquids (henceforth called {\it
  dust}).  The understanding of such object's atmospheres requires
therefore the modelling of convection and, because the
inertia of the fluid is larger than its friction (Re$\gg 10^4$),
turbulent dust formation must be considered. Hence, substellar
atmospheres involve various scale regimes with each being dominated by
possibly different physical (\eg streams, waves, precipitation) and
chemical processes (\eg combustion, dust formation, coagulation).

The large-scale structure of compact, substellar atmospheres is
characterised by local and global convective motions (\eg
thunderstorms and monsoon like winds) and -- simultaneously -- by the
gravitational settling of the dust. This scale regime has widely been
investigated by 1D static, frequency dependent atmosphere calculations
applying mixing length theory.  The presence of dust in form of
several homogeneous constituents has been modelled by applying
local stability and time scale arguments (Rossow 1978, Borrows\etal
1997, Saeger\plus Sasselow 2000, Ackermann\plus Marley 2001,
Allard\etal 2001, Tsuji 2002, Cooper\etal
2003)\nocite{ros78,bmhlg97,sesa2000,am2001,ahats2001,tsu2002a,csmlb2002}.
Recently, Woitke\plus Helling (2003a; Paper~II)\nocite{wh2003a} have
proposed a consistent treatment of nucleation, growth,
evaporation and gravitational settling of heterogeneous dust
particles, which has been applied for the first time to stellar atmosphere
models in (Woitke\plus Helling 2004; Paper~III)\nocite{wh2003b}.

Modelling macroscopic scales, Ludwig\etal (2002)\nocite{lah2002} have
presented the first 3D simulations of M-dwarf convective
atmospheres, applying the Large Eddy Simulation (LES) technique
based on the experience of Nordlund \& Stein with the solar
convection. While the largest scales, which contain most
of the energy and energise a small-scale turbulent fluid field, are
computationally resolved, smaller scales are modelled by a
hyperdiffusion which prevents the energy to accumulate in the
smallest scales, thereby smoothing out all small-scale
structures. Numerically, it stabilises the flow by filtering out
sound and fast mode waves (see \eg Caunt\plus Korpi
2001)\nocite{cako2001}.

Coming from the opposite site of the turbulent energy cascade,
Helling\etal(2001; Paper~I)\nocite{holks2001b} have investigated
the dust formation process in the small, microscopic scale regime
($l_{\rm ref}\ll H_{\rho}$) by direct simulations of acoustic wave
interactions.  These investigations of the dense, initially
dust-hostile layers in brown dwarf atmospheres have revealed a {\it
feedback loop} which characterises the dust formation process:
\begin{center}
  \parbox[h]{8cm}{The interaction of small-scale perturbations
    of the fluid field can cause a short-term temperature
    decrease low enough to initiate dust nucleation. The seed
    particles grow until they reach a size where the dust opacity is
    large enough to re-enforce radiative cooling which causes the
    temperature to decrease again below the nucleation
    threshold.  Dust nucleation is henceforth re-initiated which
    results in a further intensified radiative cooling. The nucleation
    rate and consequently also the amount of dust particles increase
    further.  This {\sc run-away} process is stopped if either the
    radiative equilibrium temperature of the gas is reached or all
    condensible material is consumed.  Meanwhile, the seed particles
    have grown to macroscopic sizes ($\mu$m).}
\end{center}
Based on this knowledge, we extent our studies to larger and larger
spatial scales aiming finally at the simulation of dust formation in
the macroscopic scale regime, i.\,e., the complete atmosphere. The
next step is therefore to study the mesoscopic scale regime ($l_{\rm
ref} < H_{\rho}$) where we model driven turbulence by stochastically
superimposed waves in the inertial Kolmogoroff range and study the
response of the dust complex. Necessary criteria are
derived for a small-scale closure model to be applied in large
scale simulations of dust forming systems.

The aim of such a scale-wise investigation is to understand the major
physical mechanisms which are responsible for the structure formation
in the atmospheres of substellar objects and to provide the necessary
informations for building an appropriate sub-grid model needed to solve
the closure problem inherent to any macroscopic turbulence simulation
(not only) of dust forming media (see also Canuto 1997a,
2000)\nocite{can97a,can2000}.  The challenge is that only the largest
scales of the turbulence cascade are comparable with real
astrophysical observations but structure formation is usually seeded
on the smallest scales especially if it is correlated with chemical
processes.  From the theoretical point of view, turbulence in thin
atmospheric layers may be of quasi-2D-nature (Cho\plus Polvani~1996,
Menou\etal 2003)\nocite{mcsh2003,copo1996}. The 2D turbulence is
characterised by an inverse energy cascade (transfer from small to
large scales), contrary to 3D turbulence, which makes a scale-wise
investigation even more urgent.

Various model approaches have been carried out to simulate and to
study turbulence in different astrophysical scale regimes.  For
example, thermonuclear flames in type Ia supernovae have been studied
on small scales by R\"opke, Niemeyer \&
Hillebrandt~(2003)\nocite{rnh2003}, investigating the Landau-Darrieus
instability in 2D simulations which is responsible for the formation
of a cellular structure of the burning front. Reinecke, Hillebrandt \&
Niemeyer~(2002)\nocite{rehn2002} have performed large-scale
calculations in order to model supernovae explosions on scales of the
stellar radius. Mac Low~(1999; see also Mac Low\plus Klessen
2004)\nocite{ml1999,mk2004}, for instance, have set up isothermal initial
velocity perturbations with an initial power spectrum of developed
turbulence in the Fourier space and initially constant density. They
model decaying turbulence on the small scales of the inertial
subrange.  Smith, Mac Low \& Heitsch~(2000)\nocite{slh2000} use a
similar approach but study the effect of driven turbulence in the same
scale regime of star-forming clouds. A stationary but stochastic
velocity field was applied by Wallin, Watson \&
Wylf~(1998)\nocite{www98} in order to perform radiative transfer
calculations of Maser spectra of a sub-parsec disk of a massive black
whole. A fundamental theoretical investigation of the methods of
driven turbulence is provided in (Eswaran \& Pope 1987,
1988)\nocite{ep88a,ep88b}. A different approach of turbulence and
convective modelling is followed by Canuto (\eg 1997b)\nocite{can97b}
who treats turbulent convection by a Reynold separation ansatz where
decomposed quantities (background field + fluctuations) are introduced
into the model equations.

In this paper, we present a model for driven turbulence which allows
us to study the onset of dust formation under strongly fluctuating
hydro- and thermodynamic conditions in the mesoscopic scale regime. We
thereby intend to model a constantly occurring energy input from the
convectively active zone.  Section~\ref{sec:ModelProblem} states the
model problem and the characteristic numbers of our astrophysical
problem. The turbulence model is outlined in
Sect.~\ref{ssec:turbmodel}. In Sect.~\ref{sec:Results}, the results
are presented for 1D simulations and an illustrative 2D example.
Section~\ref{sec:disc} contains the discussion, and
Section~\ref{sec:concl} the conclusions.

\section{The model}\label{sec:ModelProblem}

The model equations for a compact substellar atmosphere are
summarised. The model is threefold: i) hydro- and thermodynamics ({\it
Complex A}), ii) chemistry and dust formation
({\it Complex B}; both Sect.~\ref{ssec:ModelHD+Dust}), and iii) turbulence
(Sect.~\ref{ssec:turbmodel}).  Our model philosophy is to study the
dust forming system in different scale regimes in order to identify
major mechanisms which might be responsible for cloud formation and
possible variability in substellar atmospheres. The approach is based
on dimensionless equations such that their solution is characterised
by a set of characteristic numbers.

\subsection{The model for a compressible, dust-forming gas}\label{ssec:ModelHD+Dust}
The complete set of model equations was outlined in detail in Paper~I.
Only a short summary is therefore give here.

\smallskip {\it Complex A: The hydro- and thermodynamics} are
described following the classical approach for an inviscid,
compressible fluid; (Eqs.(1)--(5)) in Paper~I\footnote{Note that
  Eq.~(2) in Paper~I should be corrected as\\ $(\rho {\boldsymbol
    v})_t + \nabla\cdot (\rho {\boldsymbol v}\circ {\boldsymbol v}) =
  -\frac{1}{\rm {\gamma {\rm M^2}}} \nabla P - \frac{\rm 1}{\rm
    Fr^2}\rho {\boldsymbol g}$.}.  

\smallskip {\it Complex B: The chemistry and dust formation.} The dust
formation is a two step process -- nucleation and growth (Gail\etal
1984; Gail \& Sedlmayr~1988\nocite{gks84, gs88}) -- and depends
through the amount of condensible species on the local density
and chemical composition of the gas which are determined by Complex
A. The nucleation rate $J_*$, which is strongly
temperature-dependent, is calculated from Eq.~(17) in Paper~I
applying the modified classical nucleation theory\footnote{Classical
nucleation theory has often been criticised (\eg Michael\etal
2003)\nocite{mnl2003} but no other consistent and in hydrodynamic
simulations applicable theory for phase transitions was proposed so
far. The most accurate way is the solution of the complete chemical
rate net work for which, however, the necessary data are simply not
available. The conceptional weakness of the classical nucleation
theory is mainly the use of the bulk surface tension $\sigma$
to express the binding defects on the surface of small
clusters. However, Gail\etal(1984)\nocite{gks84} have proposed the
{\it modified} classical nucleation theory where the bulk surface
tension is {\it not} used. Instead, thermodynamic data for individual
clusters are adopted which are provided by extensive quantum
mechanical calculations (see \eg Jeong\etal 1998, 2000; Chang\etal
1998, 2001; Patzer\etal 2002; John
2003\nocite{jwfs97,cpsss01,pcjbs02,cjps98, jcss00, joh2003}).  In
contrast, because the calculation of high-quality thermodynamic data
is a challenging problem, most people base their dust formation
considerations simply on stability arguments (see \eg Tielens\etal
1998)\nocite{twmj98} which is, however, a necessary but not a sufficient
condition for phase transitions.} (Gail\etal 1984)\nocite{gks84}. The
dust growth is described by the momentum method developed by (Gail\plus
Sedlmayr 1986, 1988\nocite{gs86, gs88}; Dominik et
al.~1993\nocite{dsg93}) in combination with the differential equations
describing the element conservation (Eqs.(6)--(8) in Paper~I).

Our model of dust formation considers a prototype-like phase
transition (gas $\to$ solid) which is triggered by the nucleation of
homogeneous ${\rm (TiO_2)}_N$-clusters (see Jeong\etal 1998, Gail\plus
Sedlmayr~1998, Jeong\etal 2003)\nocite{jws98, jwbs2003, gs98a}. The
formation of the dust particles is completed by the growth of a
heterogeneous mantle which is assumed to be arbitrarily stable.
The most abundant elements after H, C, O, and N in a solar
composition gas are Mg and Si followed by Fe, S, Al, $\ldots$, Ti,
$\ldots$ Zr. Therefore, the main component of the dust mantle can
be expected to be some kind of silicate with a Mg/Si/O
mixture plus some impurities. Since the
focus of our work concerns the initiation of the dust formation in
hostile turbulent environments rather than  a detailed
description of the growth process, evaporation and drift are
neglected.  Therewith, the maximum effects regarding the amount of
dust formed in brown dwarf atmospheres is studied.

The most abundant Si bearing species in the gas phase under
conditions of chemical equilibrium is the SiO molecule.
{Therefore, the collision rate with SiO can be expected to limit the
growth of various silicate materials (like SiO$_2$,
Mg$_{2x}$Fe$_{2(1-x)}$SiO$_4$ and Mg$_{x}$Fe$_{1-x}$SiO$_3$) rather
than the collision rate with the nominal molecules (monomers) which
are usually much less stable and hence barely present in the gas phase
(Gail\plus Sedlmayr~1999\nocite{gs99}). Consequently, SiO is identified
as the key species for the description of the growth of the dust
mantles in our model (compare Paper~II). In order to prevent an
overproduction of seed particles, we furthermore include TiO$_2$
also as additional growth species, which leads to a quick consumption
of Ti from the gas phase as soon as relevant amounts of dust are present.}

\subsubsection{Characteristic numbers and scale analysis}\label{sssec:charnumb}

The use of dimensionless equations (Eqs.~(1)--(8) in Paper~I) provides
the possibility to characterise the systems behaviour by
non-dimensional numbers.  The related estimations of typical time and
length scales are summarised in Table~\ref{tab:kennz}
(Appendix~\ref{append:analysischarnumb}) for which the reference
values only need to be known by orders of magnitude. It would,
however, be very difficult to adopt an unique representation of the
reference values from recent brown dwarfs model atmosphere calculation
because of the differences among the different groups (compare
Fig.~\ref{fig:0D}). The agreement is nevertheless good enough that we
consider them as typical, classical hydrostatic brown dwarf model
atmosphere which lead our choice of reference values in
Table~\ref{tab:kennz}.

\paragraph{\underline{\it Complex A:}}

-- Assuming the typical turbulence velocity to be of the order of one
tenth of the velocity of sound leads to a {\it Mach number} $M\approx
\cal{O}$$(0.1)$.  This choice has been guided by the results of
Ludwig\etal(2002) who derived a maximum vertical velocity of
$\cal{O}$$(10^4\mbox{cm\,s$^{-1}$})\approx c_s/10$ cm\,$s^{-1}$ which
is about the same order of magnitude like the convective velocities
derived from the mixing length theory (MLT)\footnote{Model
  atmosphere calculations for Brown Dwarfs using MLT provide a typical
  (static) convective velocity $v_{\rm conv}^{\rm
    MLT}=\cal{O}$$(10^2\,\ldots\,10^3 \mbox{cm/s})\approx c_{\rm
    s}/1000\,\ldots\,c_{\rm s}/100$\,cm/s which strongly contradicts
  the value used to model an additional spectral line broadening
  component, the so-called micro-turbulence velocity $v_{\rm
    micro}\approx 1$km/s.  While $v_{\rm conv}^{\rm MLT}$ is needed to
  calculate an adequate temperature structure of the inner atmosphere,
  $v_{\rm micro}$ is needed for a best fit of the observed spectra.
  This dilemma can, however, not be solved without either a consistent
  convective theory or a direct numerical simulation (DNS) of the
  turbulence and the convective energy transfer which influences the
  local and the global temperature and velocity fields and thereby the
  observed spectral lines and their broadening.}.  This value of a
  large-scale velocity does presently determine the energy dissipation
  rate (Eq.~\ref{equ:energydiss}) of the turbulence model applied
  (Sect.~\ref{ssec:turbmodel}) in this paper.

-- The {\it Froude number} is $ Fr =\cal{O}$$(10^{-2}\ldots 10^{-1})$
for a mesoscopic reference lengths $l_{\rm ref}<H_{\rho}$.  Therefore,
the pressure gradient and the gravity are now of almost comparable
importance.  However, gravity will gain considerable influence on the
hydrodynamics only for scales regimes $l_{\rm ref}\geqslant H_{\rho}$
($l_{\rm ref}=H_{\rho} \Rightarrow Fr = M$). The analysis of the
characteristic combined drift number performed in Paper~III (see
Tables 2,~4 therein)\nocite{wh2003a} has shown that the drift term in
the dust moment equations is merely influenced by the gravity and the
bulk density of the grains. We therefore assume also in the mesoscopic
scale regime position coupling between dust and gas which seems
reasonable due to the almost equal importance of the source terms in
the equation of motion.

-- The estimate of the {\it Reynolds number}, $ Re= 10^7\ldots
10^{9}$, for a brown dwarf atmosphere situation in the mesoscopic
scale regime indicates that the viscosity of the gas is too small to
damp hydrodynamical perturbations on the largest scale to be
considered, $l_{\rm meso}$, and a turbulent hydrodynamic field can be
expected. $Re$ has increased by about one order of magnitude compared
to the microscopic scale regime (compare Table~1 in Paper~I).
Therefore, the viscosity of the gas decreases for mesoscopic scale
effects in comparison to the microscopic regime.  This is correct
since for $l_{\rm ref}=\eta \Rightarrow Re=1$ ($\eta$ - Kolmogoroff
dissipation scale) and viscosity dissipates all the turbulent kinetic
energy of the fluid.

-- In radiatively influenced environments the {\it characteristic
  number for the radiative heating / cooling}, $ Rd=4 \kappa_{\rm
  ref}\sigma T^4_{\rm ref} \cdot \frac{t_{\rm ref}}{ P_{\rm ref}}$.
The systems scaling influences $Rd$ by the reference time $t_{\rm
  ref}$ which can increases with increasing spatial scales.

\paragraph{\underline{\it Complex B:}}
The scaling of the dust moment equations provides
two {\it Damk\"ohler numbers} for dust nucleation, $Da^{\rm nuc}_{\rm d}$,
and dust growth, $Da^{gr}_{\rm d}$, and characteristic numbers for the
grain size distribution, the Sedlma{\"y}r number $Se_j$ (j - order of
dust moments, ${\rm j}\in \mathbb{N}$). Element conservation is
characterised by $ El$, the element consumption number. $Se_j$ and $
El$ are not influenced by the scaling of the system (compare
Table~\ref{tab:kennz}) but the two dust Damk\"ohler numbers, $Da^{\rm
nuc}_{\rm d}$ and $Da^{gr}_{\rm d}$, increase with increasing time
scale.

\medskip The analysis of the characteristic numbers shows that the
governing equations of our model problem are still those of an
inviscid, compressible fluid which are coupled to stiff dust moment
equations and an almost singular radiative energy relaxation if dust
is present.  The dust equations become even more stiff than in the
microscopic regime which caused severe numerical difficulties in
solving the energy equation which is coupled to the dust complex by
the absorption coefficient $\kappa$ (see
Sect.\ref{sssec:stiffcouple}). The {\it dominant interactions} occur
in the energy equation and in the dust moment equations
(Eqs.~(4),~(6),~(7) in Paper~I) also in the mesoscopic scale regime.

\subsection{The model for compressible, driven turbulence}\label{ssec:turbmodel}
A turbulent fluid field is determined by the stochastic character of
the hydrodynamic and thermodynamic quantities due to possible
interaction of different scales which are represented by inverse
wavenumbers in our model. We have constructed a pseudo-spectral method
where randomly interacting waves are generated inside a wavenumber
interval $[k_{\rm min}, k_{\rm max}]$ on an equidistant grid of $N$
wavenumbers $k_{\rm i}$ in the Fourier space. The wavenumber interval
is part of the inertial subrange of the turbulent energy cascade
(Eq.~\ref{equ:KolSpec}).

A disturbance $\delta \alpha(\boldsymbol{x},t)$ is added to a homogeneous
background field $\alpha_0(\boldsymbol{x},t)$ such that for a suitable variable
\begin{equation}
\alpha(\boldsymbol{x},t)=\alpha_0(\boldsymbol{x},t) + \delta \alpha(\boldsymbol{x},t),
\label{equ:distrubance}
\end{equation}
with $\boldsymbol{x}$ the spatial vector and $t$ the time coordinate.  The
present model for driven, compressible turbulence comprises a stochastic,
dust-free velocity, pressure and entropy field, \ie $\alpha(\boldsymbol{x},t)\, \epsilon \,
\{ \boldsymbol{u}(\boldsymbol{x},t), p(\boldsymbol{x},t), S(\boldsymbol{x},t) \}$.

\paragraph{Stochastic distribution of velocity amplitudes $\delta \boldsymbol{u}(\boldsymbol{x},t)$:}
An arbitrary scale -- represented by a wavenumber interval $k\ldots
k+dk$ ($k = \|\boldsymbol{k}\|$) -- inside the inertial range of
developed turbulence contains the energy per mass $e(k)\,dk$ in 3D, where
\begin{equation}
e(k)\,dk = C_{\rm K}\varepsilon^{2/3}k^{-5/3}.
\label{equ:KolSpec}
\end{equation}
$\varepsilon$ is the energy dissipation rate [cm$^2$/s$^3$] and
$C_{\rm K}\approx 1.5$ is the (dimensionless) Kolmogoroff constant
(see Dubois, Janberteau, Teman 1999, p.51)\nocite{djt99}.  Kolmogoroff derived this first order
description of the energy spectrum for the inertial range assuming
self-similarity of the corresponding scales. The energy spectrum
(Eq.~\ref{equ:KolSpec}) has been well verified by experiments and
simulations (see \eg Dubois, Janberteau, Teman 1999)\nocite{djt99}.
In the framework of Kolmogoroffs theory, $\varepsilon$ is constant
for all scales $k$ and time $t$ ({\it homogeneous, isotropic
  turbulence}).

The energy dissipation rate can be estimated if already one typical
scale and its corresponding reference velocity is known because
$\varepsilon$ is assumed to be constant for all scales (see also
Sect.~\ref{sssec:charnumb} {\it Complex A}). From dimensional arguments,
\begin{equation}
\label{equ:energydiss}
\varepsilon = C_1  \frac{u^3}{l} = C_1 \frac{u(k_{\rm min})^3}{l(k_{\rm min})},
\end{equation}
for instance for the largest scales of interest inside the inertial
range, \ie for the smallest wavenumber $k_{\rm min}$. According to
Jimenez\etal(1993)\nocite{jwsr93}, $C_1=0.7$.

A wavenumber interval [$k_{\rm i}, k_{\rm i+1}$] contains, according
to Eq.~(\ref{equ:KolSpec}), the turbulent kinetic energy density per
mass $E^{\rm i}_{\rm turb}$, 
\begin{equation}
E^{\rm i}_{\rm turb} =\int_{k_{\rm i+1}}^{k_{\rm i}} e(k)\,dk
         = \frac{3}{2}C_{\rm K}\varepsilon^{2/3}[k_{\rm i}^{-2/3} - k_{\rm i+1}^{-2/3}].
\label{equ:Eturb}
\end{equation}
The square of the velocity amplitude, $A_{\rm u}(\bar{k_{\rm i}})$, is
correlated with the turbulent kinetic energy in Fourier space by
\begin{equation}
A_{\rm  u} (\bar{k_{\rm i}}) = \sqrt{ 2 z_3 E^{\rm i}_{\rm turb}},
\label{equ:deltauFour}
\end{equation}
with
\begin{equation}
\bar{k_{\rm i}} = \frac{k_{\rm i} +k_{\rm i+1}} {2}
\label{equ:deltak}
\end{equation}
the mean value of $k$ in the wavenumber interval considered.  $k_{\rm
  i}$ are $N$ equidistantly distributed wavenumbers in the Fourier
space, (i$=1, \ldots N$) with $N$ the number of modes. The $k_{\rm i}$
are chosen between $k_{\rm min}$ and $k_{\rm max}$ when the
calculation is started and are kept constant further on.

Here, the so-called {\it ultraviolet truncation} $A_{\rm u}(k_{\rm
  i})=0$ for $k_{\rm i} > k_{\rm max}$ is applied in order to avoid
the infinite energy problem of the classical field theories in
Eq.~(\ref{equ:Eturb}) (stated in Bohr 1998, p.23)\nocite{??10}. The
minimum wavenumber is determined by the largest scale $l_{\rm ref}$,
\ie the size of test volume.  Only wavenumbers inside a sphere of
radius $k_{\rm max}$ excluding the origin are forced (see also
Overholt \& Pope 1998, p.13)\nocite{op98}.

Assuming the ergodic hypothesis (see \eg Frisch
1995)\nocite{frisch95}, the turbulent kinetic energy $E_{\rm
  turb}^{\rm i}$ (Eq.~\ref{equ:Eturb}) was assumed to be the most
likely value (compare \eg Mac Low\etal 1999) with a stochastic
fluctuation generated by a zero-centred Gaussian distributed random
number $z_3>0$ according to the Box-M\"uller formula
\begin{equation}
z_3 = \sqrt{-2\,\log z_1} \sin(\pi z_2),
\label{equ:BoxMueller}
\end{equation}
$z_1$ and $z_2$ are equally distributed random numbers $\,\epsilon\, [0, 1)$.

A cosine Fourier transformation provides the real values of the
velocity amplitude $\delta
\boldsymbol{u}(\boldsymbol{x},t)$ in ordinary space,
\begin{equation}
\delta\boldsymbol{u}(\boldsymbol{x},t) = \Sigma_{\rm i}\, A_{\rm u}(\bar{k_{\rm i}})
\cos(\bar{\boldsymbol{k_{\rm i}}}\boldsymbol{x} - \omega_{\rm i}t + \varphi_{\rm i})
\hat{\bar{\boldsymbol{k}}},
\label{equ:deltauRaum}
\end{equation}
with $\bar{\boldsymbol{k_{\rm i}}}=\bar{k_{\rm
    i}}\hat{\bar{\boldsymbol{k}}}$ and
$\boldsymbol{u}(\boldsymbol{x},t)=u(\boldsymbol{x},t)\hat{\boldsymbol{u}}$.
Also the directions of
$\boldsymbol{k_{\rm i}}$, \ie the direction of
$\delta\boldsymbol{u}(\boldsymbol{ x},t)$, are chosen randomly
according to
\begin{eqnarray}
\label{equ:randomk1}
\hat{\bar{k}}_{\rm i, x} = \sin\alpha\,\cos\beta & \qquad & \cos\alpha=1-2\,z_4\\
\label{equ:randomk2}
\hat{\bar{k}}_{\rm i, y} = \sin\alpha\,\sin\beta & \qquad & \sin\alpha = \sqrt{1.0-(\cos\alpha)^2}\\
\label{equ:randomk3}
\hat{\bar{k}}_{\rm i, z} =  \cos\alpha  & \qquad & \beta =  2\pi\,z_5,
\end{eqnarray}
with $z_4$ and $z_5$ equally distributed random numbers.  The 1D and
2D case of Eq.~(\ref{equ:deltauRaum}) is obtained by projection. A
longitudinal wave results in 1D.

$\varphi_{\rm i}=2\pi\,z_6$ is the equally distributed random phase
shift which is chosen separately for each wavenumber.
$\omega_{\rm i}$ is the angular velocity for which a {\it dispersion
  relation} is derived from dimensional arguments. It follows from
Eq.~(\ref{equ:energydiss}) that for each scale $l_{\rm i} =
2\pi/\bar{k}_{\rm i}$ the corresponding eddy turnover time $t_{\rm i}$ results to be
\begin{equation}
u_{\rm i} \sim (\varepsilon\, l_{\rm i})^{1/3} 
\,\Rightarrow\, 
t_{\rm i} \sim \left( \frac{\varepsilon}{l_{\rm i}^2}\right)^{-1/3}.
\label{equ:uiti}
\end{equation}
Since per definition $\omega_{\rm i}=2\pi/t_{\rm i}$ the {\it
dispersion relation in the inertial subrange} is
\begin{equation}
\omega_{\rm i} = (2\pi\,\bar{k}_{\rm i}^2\varepsilon)^{1/3}.
\label{equ:disrel}
\end{equation}

\paragraph{Stochastic distribution of pressure amplitudes $\delta p(\boldsymbol{x},t)$:} 

The pressure amplitude is determined depending on the wavenumber of
the velocity amplitude $A_{\rm u}(\bar{k_{\rm i}})$ such that the
compressible (sound waves) and the incompressible pressure limits are
matched for the smallest $k_{\rm min}$ and the largest $k_{\rm max}$
wavenumber, respectively,
\begin{equation}
A_{\rm p}(k_{\rm i}) = -\frac{[k_{\rm max} - k_{\rm i}]\, \rho A_{\rm u}(\bar{k_{\rm
    i}})^2 + [k_{\rm i} - k_{\rm min}]\, \rho  c_{\rm s} A_{\rm u}(\bar{k_{\rm i}})}{[k_{\rm max} -  k_{\rm min}]}.
\label{equ:deltapFou}
\end{equation}
The maximum wavenumber $k_{\rm max}=2\pi/(3\,\Delta x)$ is determined by some
factor (here 3; see Overholt \& Pope 1998 for discussion)\nocite{op98}
of the spatial grid resolution $\Delta x$ (see Table~\ref{tab:kennzalle}).

A spectral decomposition (compare Eq.~\ref{equ:deltauRaum}) provides the real values of the
pressure amplitude $\delta p(\boldsymbol{x},t)$  in ordinary space,
\begin{equation}
\delta p (\boldsymbol{x},t) = - \Sigma_{\rm i}\, A_{\rm p}(\bar{k_{\rm i}})
\cos(\bar{\boldsymbol{k_{\rm i}}}\boldsymbol{x} - \omega_{\rm i}t + \varphi_{\rm i}).
\label{equ:deltapRaum}
\end{equation}

\paragraph{Stochastic distribution  of the entropy $S(\boldsymbol{x},t)$:}

The entropy $S(\boldsymbol{x},t)$ is a purely thermodynamic quantity
and a distribution can in principle be chosen independently from the
distribution of the hydrodynamic quantities.  In the adiabatic case,
$S(\boldsymbol{x},t)$ is conserved along particle trajectories. So
far, $S(\boldsymbol{x},t)$ has been kept constant.

For a given $S(\boldsymbol{x},t)$ and $p(\boldsymbol{x},t)$ (see
Eq.~\ref{equ:distrubance}) the gas temperature $T(\boldsymbol{x},t)$
is given by
\begin{equation}
\log T(\boldsymbol{x},t) = \frac{S(\boldsymbol{x},t) + R\log p(\boldsymbol{x},t) - R\log R}{c_{\rm V} + R},
\label{equ:Temp}
\end{equation}
with $R$ the ideal gas constant, $c_V$ specific heat capacity for
constant gas volume.

\medskip In this work, we have simulated the turbulence by {\it
  prescribing boundary conditions} (Sect.~\ref{sec:Numerics}) applying
  Eqs.~(\ref{equ:deltauRaum}),~(\ref{equ:deltapRaum}),
  and~(\ref{equ:Temp}).  Stochastically created and superposed waves
  continuously enter the model volume and are advectivelly transported
  inward by solving the model equations
  (Sect.~\ref{ssec:ModelHD+Dust}). A hydrodynamically and
  thermodynamically fluctuating field is generated which influences
  the local dust formation because it sensitively depends on the local
  temperature and density.

\begin{figure*}
\begin{center}
\epsfig{file=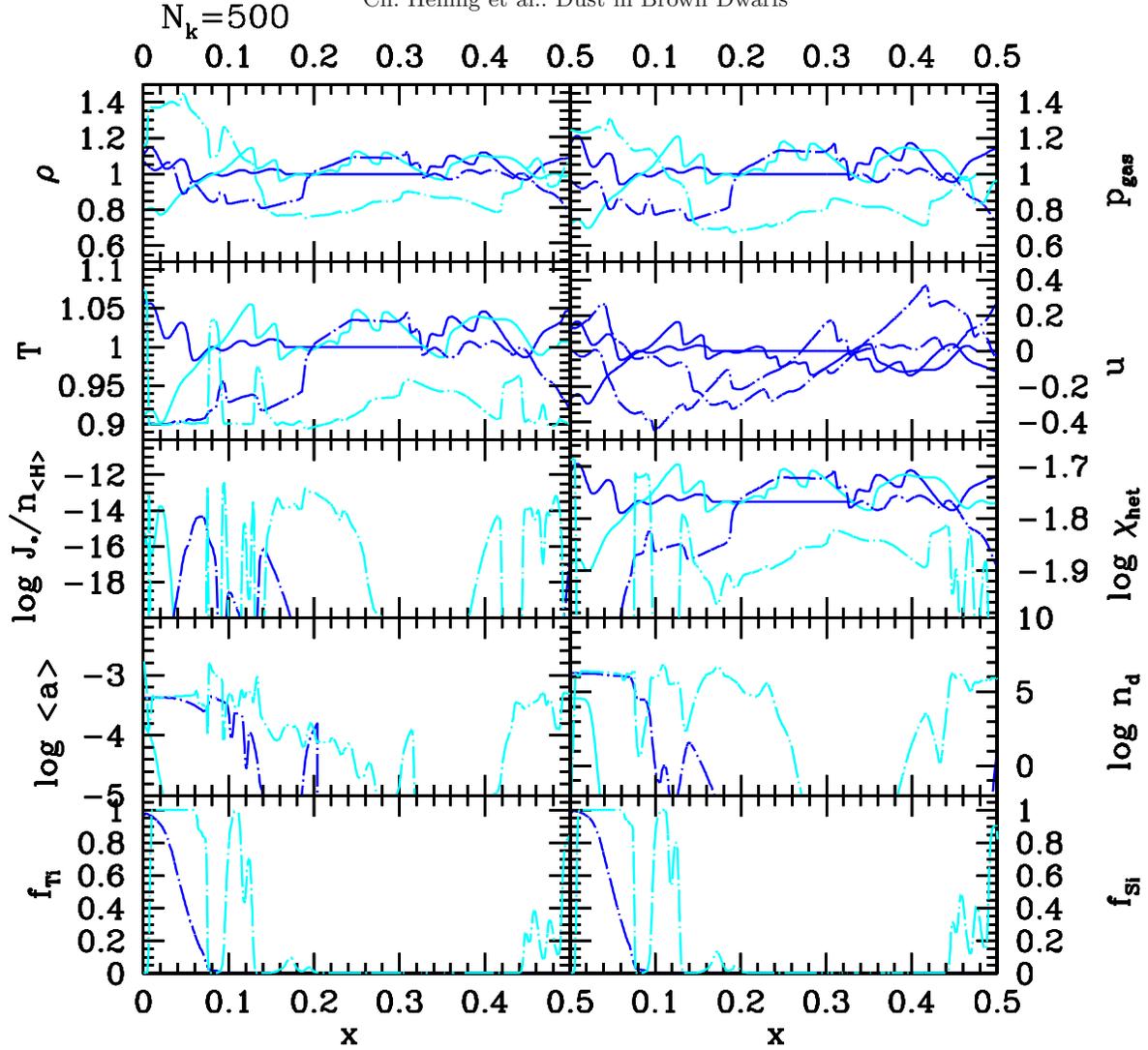, scale=0.9}
\end{center}
\caption[]{Time sequence of 4 time steps for a 1D simulation with
  $T_{\rm ref}=2100$\,K, $M=0.1$, $N_{\rm k}=500 (\,\,\nearrow\,\,$
  entree A in Table~\ref{tab:kennzalle}; black/blue solid -- 0.48s,
  grey/cyan solid -- 0.68s, black/blue dash-dot -- 1.12s, grey/cyan
  dashed-dot -- 1.5s). The first instant of time shows the
  superimposed waves which just enter the test volume. The later times
  show nucleation fronts ($t=0.68$s) and nucleation events ($t>0.5$s)
  occurring.\\ $T$, $\rho$, $u$, $p_{\rm gas}$ are given
  dimensionless, the dust quantities have their physical units 
  ($J_*/n_{\rm <H>}$ in [s$^{-1}$], $\chi_{\rm het}$ in [cm s$^{-1}$],
  $\langle a\rangle$ in [cm], $n_{\rm d}$ in [cm$^{-3}$], $f_{\rm Ti}$
  in [$-$], $f_{\rm Si}$ in [$-$])}.
\label{fig:Nk500_2100K_Ort}
\end{figure*}

\begin{figure*}
\begin{tabular}{cc}
\hspace*{-1.0cm}
\epsfig{file=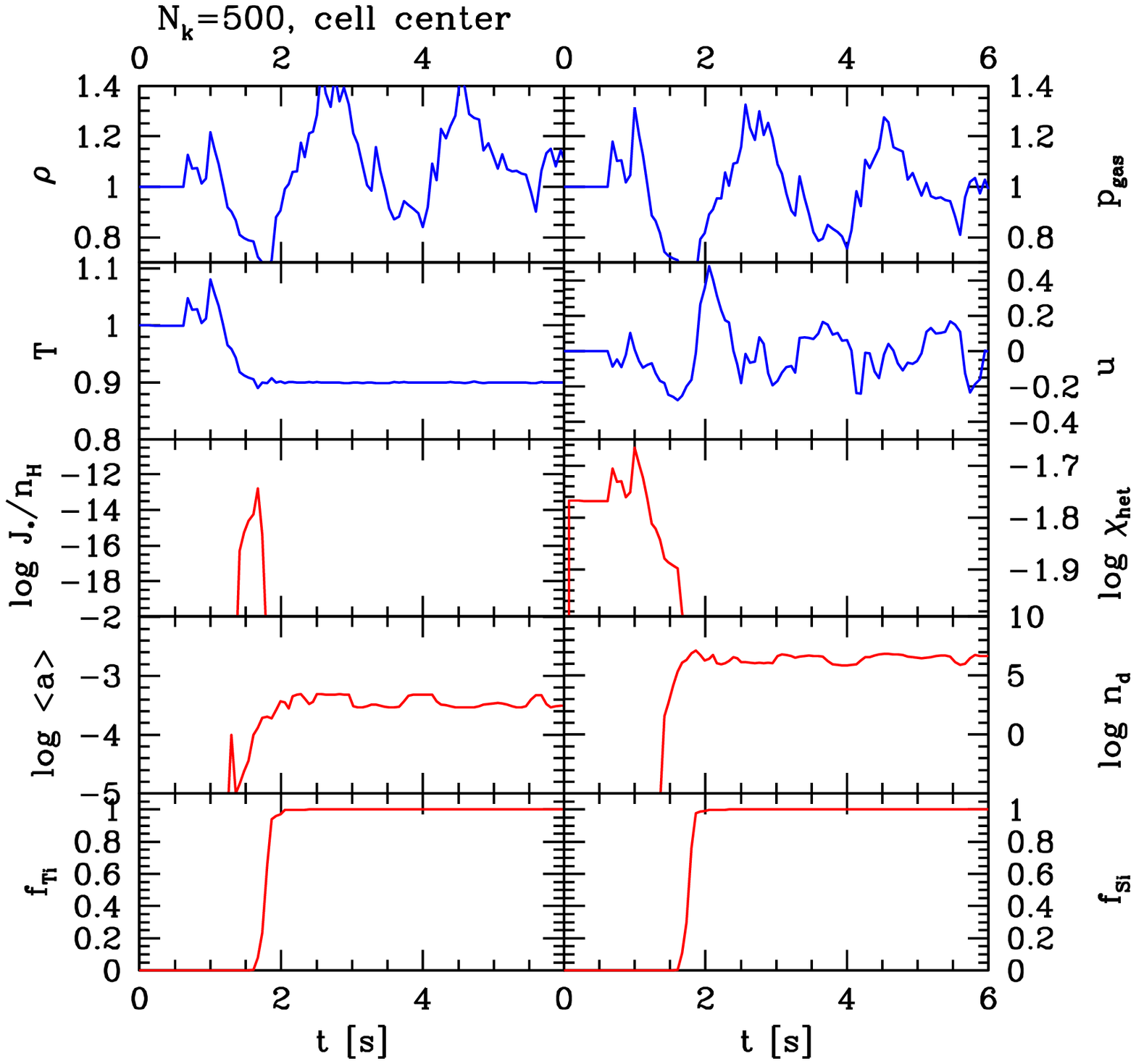,
scale=0.55}
&
\hspace*{-0.7cm}\epsfig{file=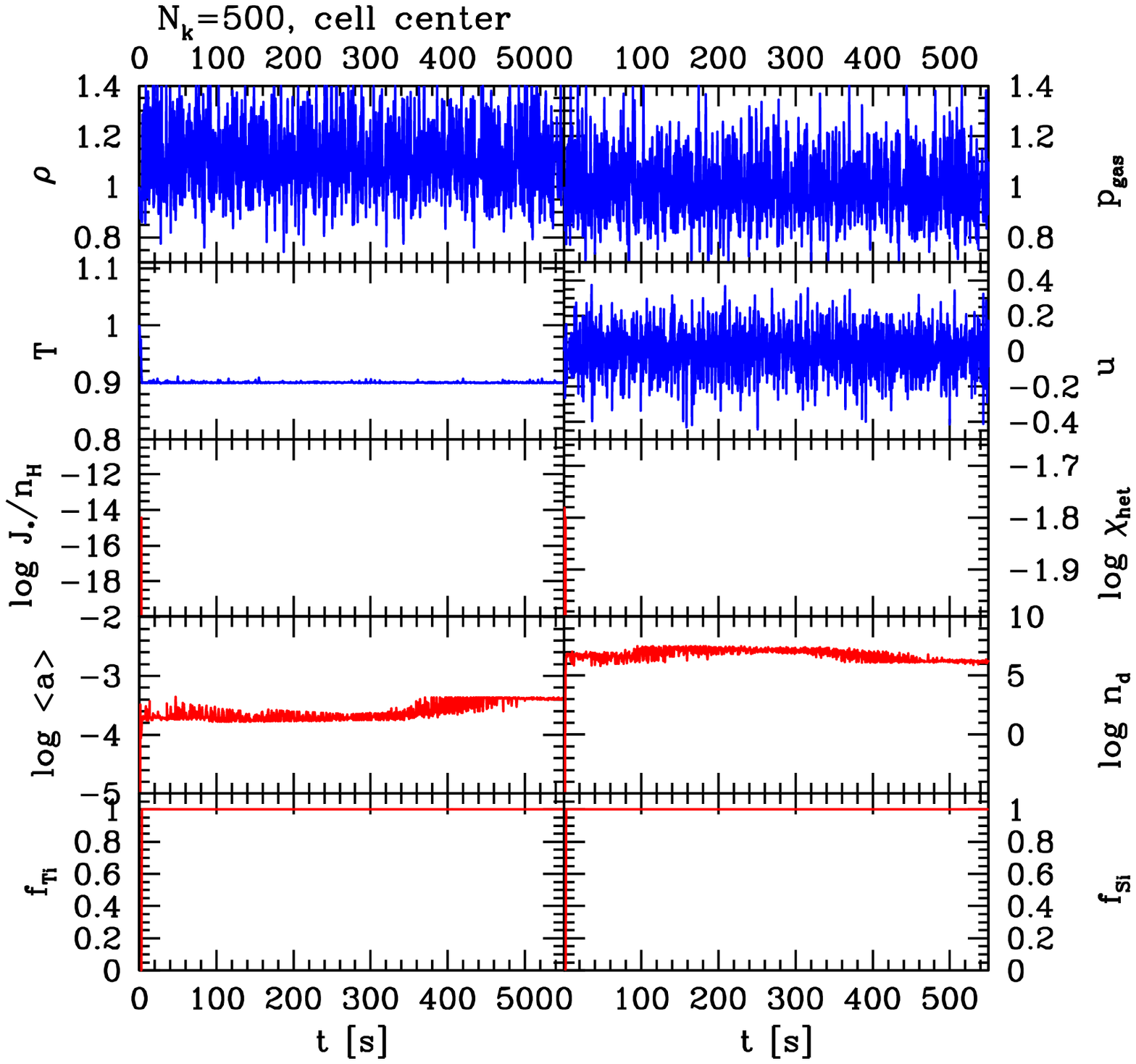,
scale=0.55}
\end{tabular}
\caption[]{Time evolution in the cell centre ($T_{\rm ref}=2100$\,K,
  $M=0.1$, $N_{\rm k}=500\,\,\nearrow\,\,$ entree A in
  Table~\ref{tab:kennzalle}).\\ {\bf Left:} During the first 6\,s
  (each 100th time step plotted). {\bf Right:} Over 550\,s\,$\approx
  9$\,min (each 1000th time step plotted)\\ The units are the same
  like in Fig.~\ref{fig:Nk500_2100K_Ort}.}
\label{fig:Nk500_2100K_zeit}
\end{figure*}

\subsection{Numerics}\label{sec:Numerics}

The fully time-dependent solution of the model equations has been
obtained by applying a multi-dimensional hydro code (Smiljanovski et
al.~1997)\nocite{smk97} which has been extended in order to treat the
complex of dust formation and elemental conservation (Eqs.~(1)--(8),
Paper~I).  The hydro code has already been the subject of several
tests and studies in computational science (see also Schneider et
al.~1998\nocite{sbgk98}, Schmidt \& Klein 2002\nocite{sk2003}).

\paragraph{Boundary conditions and turbulence driving:} 
The Cartesian grid is divided in the cells of the test volume (inside)
and the ghost cells which surround the test volume (outside). The
state of each ghost cell is prescribed by our adiabatic model of
driven turbulence (Sect.~\ref{ssec:turbmodel}) for each time. Hence,
the actual fluctuation amplitudes of the fluid field
($\delta\boldsymbol{u}(\boldsymbol{x},t)$, $\delta
p(\boldsymbol{x},t)$, $\delta S(\boldsymbol{x},t)\,\Rightarrow\,$
$\delta T(\boldsymbol{x},t)$, $\delta (\rho \boldsymbol{x},t)$) is the
result of a spectral composition of a number of Fourier modes which
are determined by the Kolmogoroff spectrum. The absolute level of this
energy distribution function is given by the velocity ascribed to the
largest, \ie the energy containing scale of the simulation (see {\it
Complex A} in Sect.~\ref{sssec:charnumb}).

The hydro code solves the model equation in each cell (test volume +
ghost cells) and the prescribed fluctuations in the ghost cells are
transported into the test volume by the nature of the HD equations.
The numerical boundary occurs between the ghost cells and the
initially homogeneous test volume and are determined by the solution
of the Riemann problem. Material can flow into the test volume and can
leave the test volume. The solution of the model problem is considered
inside the test volume.

\paragraph{Initial conditions:}
The (dimensionless) initial conditions have been chosen as
homogeneous, static, adiabatic, and dust free, \ie $\rho_0=1$,
$p_0=1$, $u_0=0$, $L_0=0\,(\Rightarrow\, L_j=0)$ in order to represent
a (semi-)static, dust-hostile part of the substellar atmosphere. This
allows us to study the influence of our variable boundaries on the
evolution of the dust complex without a possible intersection with the
initial conditions.

\subsubsection{Stiff coupling of dust and\\  radiative heating / cooling}\label{sssec:stiffcouple}

Dust formation occurs on much shorter time scales than the
hydrodynamic processes (see \eg Sect.~2.2 in
Paper~I\nocite{holks2001b}). Approaching regimes of larger and larger
scales makes this problem more and more crucial
(Sect.~\ref{sssec:charnumb}).  Therefore, the dust moment and element
conservation equations ({\it Complex B}) are solved applying an ODE
solver in the framework of the operator splitting method assuming $T,
\rho=$const during ODE solution. In Paper~I we have used the CVODE
solver (Cohen \& Hindmarsh 2000; LLNL) which turned out to
be insufficient for the mesoscopic scale regime which we attack in the
present paper.  CVODE failed to solve our model equations after the
dust had reached its steady state (compare Fig.~3 in Paper I).
Therefore, it was not possible to simulate the equilibrium situation
of the dust complex in the mesoscopic scale regime by using CVODE
which in other situation has been very efficient.

\paragraph{The LIMEX solver:}\label{sec:LIMEX}
The solution of the equilibrium situation of the dust complex is
essential for our investigation since it describes the static case of
{\it Complex B} when no further dust formation takes place (\ie where
the source terms in Eqs.~(6)--(8) in Paper I vanish). The reason may
be that all available gaseous material has been consumed and the
supersaturation rate $S=1$ or the thermodynamic conditions do not
allow the formation of dust. The first case involves an asymptotic
approach of the gaseous number density (or element abundance; see
Eqs.~8 in Paper~I) of $S=1$ which often is difficult to be solved by
an ODE solver due to the choice of too large time steps. However, the
asymptotic behaviour is influenced by the temperature evolution of the
gas/dust mixture which in our model is influenced by radiative
heating/cooling (see r.h.s. of Eq.~(3) in Paper~I). Since the
radiative heating/cooling ($Q_{\rm rad} = Rd\,\kappa(T^4_{\rm RE} -
T^4)$ heating/cooling rate) depends on the absorption coefficient
$\kappa$ of the gas/dust mixture which strongly changes if dust forms.
Consequently, the radiative heating/cooling rate is strongly coupled
to the dust complex which in turn depends sensitively on the local
temperature which is influenced by the radiative heating/cooling.  It
was therefore necessary to include also the radiative heating/cooling
source term in the separate ODE treatment for which we adopted the
LIMEX DAE solver. 

LIMEX (Deuflhard\& Nowak~1987)\nocite{dw87} is a solver for linearly
implicit systems of differential algebraic equations. It is an
extrapolation method based on a linearly implicit Euler discretisation
and is equipped with a sophisticated order and step-size control
(Deuflhard~1983)\nocite{deu83}. In contrast to the widely used
multi-step methods, \eg OVODE, only linear systems of equations and no
non-linear systems have to be solved internally. Various methods for
linear system solution are incorporated, \eg full and band mode,
general sparse direct mode and iterative solution with
preconditioning. The method has shown to be very efficient and robust
in several fields of challenging applications in numerical (Nowak\etal 1998,
Ehrig\etal 1999) and astrophysical science (Straka~2002).
\nocite{neo98,enod99,stra2002}

\section{Results}\label{sec:Results}

The simulations presented in the following are characterised by the
reference parameter set or the set of dimensionless numbers given in
Table~\ref{tab:kennzalle} (Appendix~\ref{append:analysischarnumb}),
and are carried out with an spatial resolution of $N_{\rm x}=500$ if
not stated differently.  After a detailed investigations of our 1D
models, the mean behaviour of the dust forming system is studied
(Sect.~\ref{subsec:means}) which might, nevertheless, be an easier
link to observations.  The mean values do, furthermore, provide a
first insight regarding significant features of our dust forming
system which a sub-grid model for a follow up large-scale simulation
should reproduce.  Turbulent fluctuations are discussed in terms of
apparent standard deviations.  Section~\ref{subsec:dustwindow} will
demonstrate the existence of a stochastic and a deterministic dust
formation regime in turbulent environments, beside a regime where dust
formation is impossible, \ie the problem of the dust formation window
is discussed for substellar atmospheres.

Section~\ref{subsec:2D} will illustrate how stochastically
superimposed waves trigger the dust formation process in 2D.
Large-scale (inside the mesoscopic regime) hydrodynamic motions seem
to gather the dust in larger and larger structures which is a result
of multi-dimensionality. The 1D simulations provide, however, the tool
to gain detailed insight into the interactions of chemistry and
physics for which multi-dimensional simulations are far to complex.

\begin{figure*}
\begin{tabular}{cc}
\hspace*{-1.0cm}
\epsfig{file=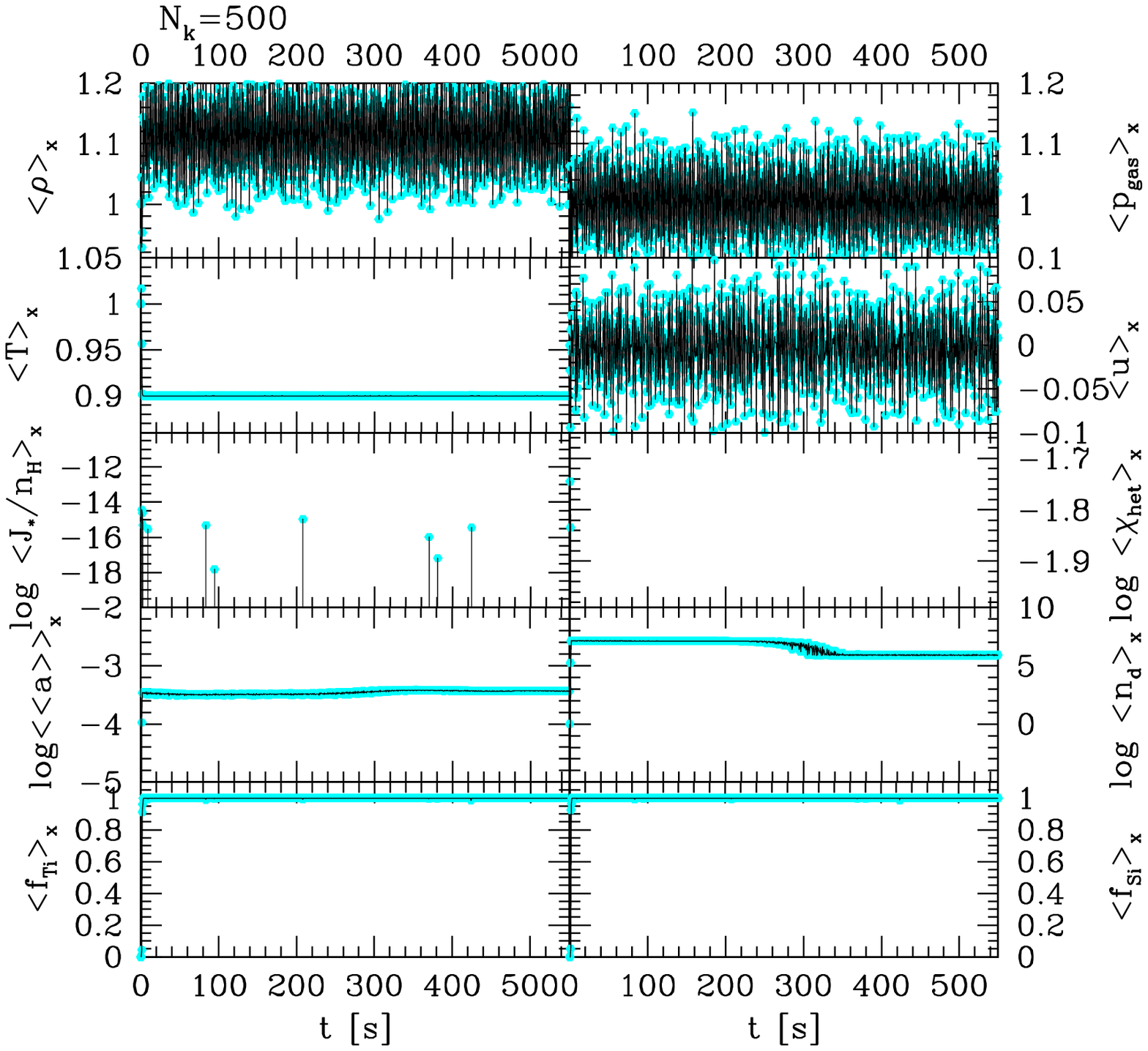, scale=0.55}
&
\hspace*{-0.7cm}
\epsfig{file=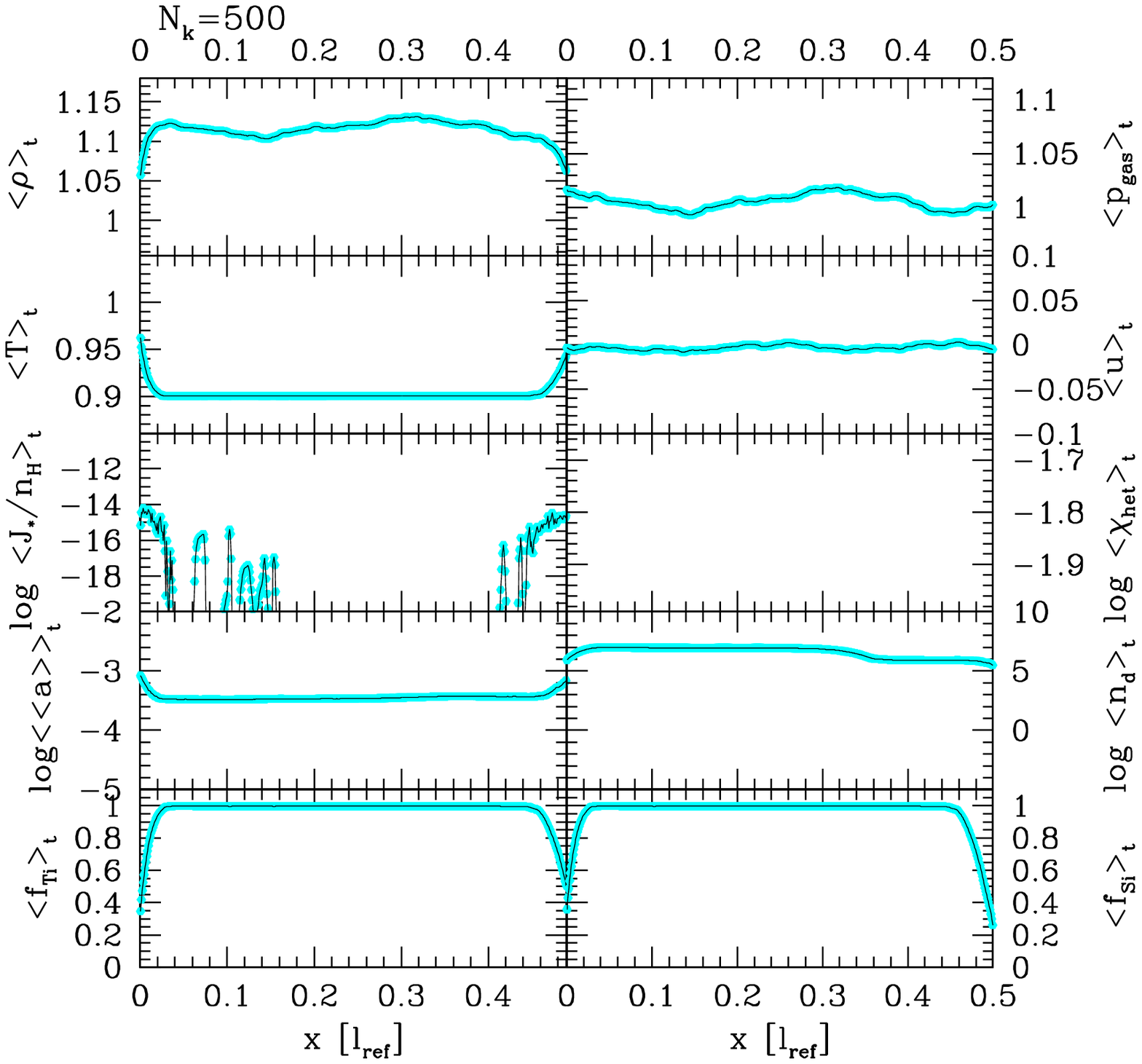,
scale=0.55}
\end{tabular}
\caption[]{Mean values ($T_{\rm ref}=2100$\,K, $M=0.1$, $N_{\rm k}=500\,\,\nearrow\,\,$ entree A in Table~\ref{tab:kennzalle}).\\  {\bf Left}: Time evolution of the space means. {\bf Right:} Time means as function of site over $10^7$ time steps.}
\label{fig:Means}
\end{figure*}

\subsection{Short term evolution}\label{subsec:shortterm}

An inviscid, astrophysical test fluid in 1D with $T_{\rm ref}=2100$\,K
and $M=0.1$ ($\nearrow$ entree A Table~\ref{tab:kennzalle}) is excited
in the wavenumber interval $[k_{\rm min}, k_{\rm max}]$ by 500 modes,
\ie  $N_k=500$, and its short term evolution is demonstrated
(Figs.~\ref{fig:Nk500_2100K_Ort},~\ref{fig:Nk500_2100K_zeit}; l.h.s.).
The smallest eddy has than a size of $\lambda^{\rm 1D}_{\rm min}=
1$\,m.  The simulations assume a 1D test volume in horizontal
direction and gravity does therefore not influence our 1D results.

\paragraph{Spatial evolution:}

Stochastically created waves move into the 1D test volume from both
sides ($t=0.48$\,s, Fig.~\ref{fig:Nk500_2100K_Ort}) with a maximum
velocity amplitude of $\cal{O}$$(10^3\mbox{cm \,s$^{-1}$})$
representing the turbulent velocity fluctuations.  At some instant of
time, the temperature disturbance due to inward moving superimposed
waves is large enough that the nucleation threshold temperature is
crossed locally ($T<T_S$, Paper~I below Eq.~(21); $t=0.68$\,s
Fig.~\ref{fig:Nk500_2100K_Ort}, grey/cyan solid line).  As the
temperature disturbance penetrates into the test volume, a {\it
nucleation front} forms which moves into the dust free gas of the test
volume and leaves behind dust seeds which can grow to considerable
sizes (compare the change of $\log n_d$ from $t=0.68$\,s (grey/cyan
solid) and $t=1.12$\,s (black dash-dot ) between $x=0$ and $x=0.16$
Fig.~\ref{fig:Nk500_2100K_Ort}).

The superimposed waves which enter the test volume through its
boundaries will also interact with each other after some time. An {\it
  event-like nucleation} results ($t=1.12$\,s, $x=0.14$
Fig.~\ref{fig:Nk500_2100K_Ort}). More dust is formed and meanwhile,
the particles are large enough to re-initiate nucleation by efficient
radiative cooling due to the strongly increased opacity ($t=1.5$\,s
Fig.~\ref{fig:Nk500_2100K_Ort}, grey/cyan dash-dot).

The result is a very inhomogeneously fluctuating distribution in size,
number and degree of condensation of dust in the test volume when the
dust formation dominated the dynamics of the system.  The fluctuations
are stronger in the beginning of the simulations and homogenise with
time (Fig.~\ref{fig:Nk500_2100K_zeit} l.h.s.). The long term behaviour will be
discussed in Sect.~\ref{subsec:xM}.

\paragraph{Time evolution:}
For better understanding of the time evolution of the hydrodynamic and
dust quantities in a stochastically excited medium, the time evolution
in the test volume's centre ({\it cell centre}) is depicted in Fig.~\ref{fig:Nk500_2100K_zeit} (l.h.s) for
the first 6s of the reference simulation with $N_k=500$. 

The dust complex reaches a steady state after about $t\approx 1.8$\,s
in the centre of the test volume in the present simulation, for which
only one singular nucleation event has been responsible (3rd panel,
Figs.~\ref{fig:Nk500_2100K_zeit} l.h.s.).  Consequently, $f_{\rm Ti} =
f_{\rm Si}=1$. In contrast, the hydro- and thermodynamic quantities
(gas density, $\rho$, gas pressure, $p$, and velocity, $u$) continue
to fluctuate considerably around their initially homogeneous values.
The small variations of the mean particle size $\langle a\rangle$ and
the number of dust particles $n_{\rm d}$ are partly caused by the
hydrodynamic motion in and out of the cell centre of the dust forming
material and partly by the turbulent fluctuations themselves.  The
thermodynamic behaviour of the dust changes from adiabatic to isotherm
as result of the strong radiative cooling by the dust. Therefore, the
temperature drops and reaches the radiative equilibrium level
($T=T_{\rm RE}$).  Exactly the same qualitative behaviour was observed
from Fig.~3 in Paper~I.

{\it We conclude a different distribution of dust inside an initially
  dust free gas element: While in the centre of a gas element the dust
  formation process is completed ($f=1$) after it was initiated by
  waves which are emitted by its surrounding, disturbances from the
  boundary prevent the boundary layers of the gas elements to reach
  $f=1$. Consequently, a convectively ascending initially dust free
  cloud can be excited to form dust by waves running through it.
  Therefore, a cloud can be fully condensed much earlier than by any
  classical, static model predicted.}

\subsection{Long term evolution}\label{subsec:xM}

The long term behaviour of our dust forming system sets in after the
dust formation process is complete ($f=1$) and radiative equilibrium
($T=T_{\rm RE}$) is reached. Due to the strong cooling capability of
the dust, only small deviations occur from the radiative equilibrium
if compression waves occur which may be seen as colliding small-scale
turbulence elements.  The general change of the temperature
$T\rightarrow T_{\rm RE}$ caused an increase of the density in the test
volume (density level $\rho > 1$ Fig.~\ref{fig:Nk500_2100K_zeit} r.h.s.)
in order to maintain pressure equilibrium (pressure level at $p\approx
1$, Fig.~\ref{fig:Nk500_2100K_zeit} r.h.s.).

The long term behaviour of $\rho$, $p$, and $u$ are characterised by
strong fluctuations constantly generated by our turbulence driving. In
Fig.~\ref{fig:Nk500_2100K_zeit} (l.h.s), which depicts a higher time
resolution of Fig.~\ref{fig:Nk500_2100K_zeit} (r.h.s.), single waves
(turbulence elements) are still distinguishable of which only spikes
out of a jungle of noise are left to observe in the long term
behaviour in Fig.~\ref{fig:Nk500_2100K_zeit} (r.h.s.).

Comparable small fluctuations of the mean particles size, $\langle
a\rangle$, and the number of dust particles, $n_d$, occur over a long
time. We recover here the 20\% fluctuation which was already
observable in Fig.~\ref{fig:Nk500_2100K_zeit} (l.h.s). Since the dust
formation process is complete, these fluctuations must be of
hydrodynamic origin, \ie caused by the movement of the small-scale
turbulence elements.

\subsection{The mean behaviour in space and time}\label{subsec:means}
The mean behaviour of a turbulent, dust forming gas is studied.  The
space mean is the average over the test volume at each time step
\begin{equation}
\label{eq:spacemean}
\langle \alpha \rangle_x(t) = \frac{1}{N_x}\sum_{i=0}^{N_x} \alpha_i(x_i, t)
\end{equation}
and  the time mean is the mean of each mesh cells over time,
\begin{equation}
\label{eq:timemean}
\langle \alpha \rangle_t(x) = \frac{1}{N_t}\sum_{i=0}^{N_t} \alpha_i(x, t_i).
\end{equation}
Both represent the most plausible values of the quantity
$\alpha(x, t)$ i) at a certain instant of time
(Eq.~\ref{eq:spacemean}), and ii) at a certain site in the test volume
(Eq.~\ref{eq:timemean}). The space means are calculated by leaving out
the cells close to the boundary in order to exclude the fluctuations in
the dust quantities due to inflowing dust-free material.
Figure~\ref{fig:Means} (l.h.s.)  depicts the space means $\langle\,
\cdot \,\rangle_{\rm x}$ as function of time $t$ [s], and
Fig.~\ref{fig:Means} (r.h.s.) depicts the time means $\langle\, \cdot
\,\rangle_{\rm t}$ as function of $x$ space [$l_{\rm ref}$].

Figures~\ref{fig:Means} shows that the space and the time mean values
differ considerably for the hydrodynamic quantities: Strong
fluctuations of the space means (l.h.s.) occur as function of time
while the time means (r.h.s.) exhibit comparatively smooth variation.
These fluctuations increase with increasing number of excitation modes
which shows that the fluctuations are of hydrodynamic origin (see also
Sect.~\ref{subsec:depNk}).

\begin{figure}
\hspace*{-0.5cm}
\epsfig{file=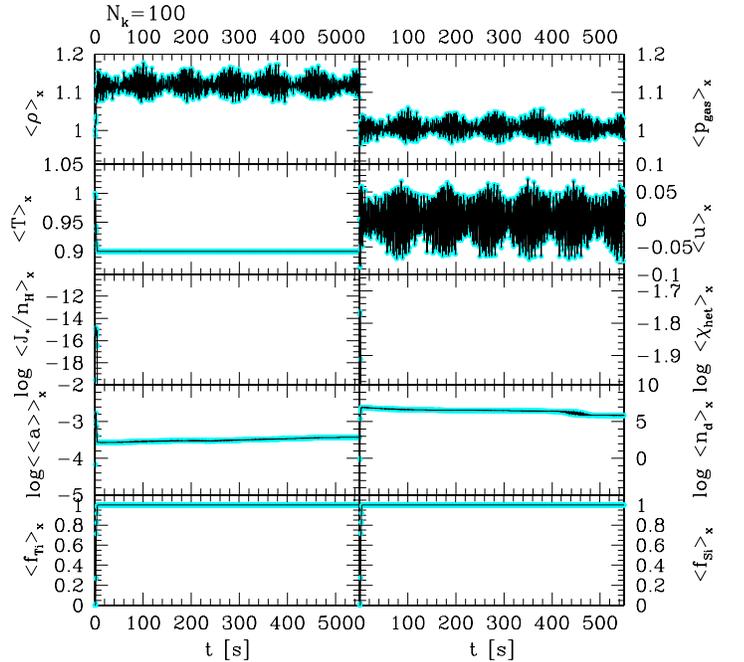, scale=0.54}
\caption[]{Same like Fig.~\ref{fig:Means} (l.h.s.) but  $N_k=100$.}
\label{fig:Nk100_2100K_spacemean}
\end{figure}

\paragraph{Space means:} 

The study of the long-term behaviour of the space means (l.h.s.,
Figs.~\ref{fig:Means}) discloses a considerable variation of the
hydrodynamic mean quantities. In contrast, the dust quantities are
almost constant in time after the dust formation process is completed.
This result is a consequence of the assumed symmetry (1D), where
every wave has to cross the whole test volume which is not the case in
a multi-dimensional fluid field (compare Sect.~\ref{subsec:2D}).

The formation of dust causes the temperature to change
considerably towards the radiation equilibrium level causing thus
the density level to change (\eg  increase if $T$ decreases) in order
to recover the pressure equilibrium. Therefore, initially small
perturbations in a dust forming system have a large effect on its
overall hydrodynamic structure.

The strong variation of the dust quantities during the beginning of
the simulation is smeared out with increasing averaging time.
Therefore, observing a spatially unresolved dust forming system over a
long time will not unhide the inhomogeneous behaviour though it will
have a profound influence of the large-scale structure of any dust
forming system, e.g., by the transition adiabatic $\rightarrow$
isothermal behaviour, by backwarming in an substellar atmosphere, or by
the enrichment and the depletion by gravitational settling.

\paragraph{Time means:}
The variation of the hydrodynamic mean values is less strong in space
(r.h.s., Figs.~\ref{fig:Means}) than in time (l.h.s.,
Figs.~\ref{fig:Means}) and resembles more common expectations for such
average quantities than the hydrodynamic space means do.  The density
shift $\rho(t=0)\,\rightarrow\,\rho(t(T=T_{\rm RE}))$ due to
$T\,\rightarrow\,T_{\rm RE}$ is easier observable than for the space
mean.

The time mean of the nucleation rate, however, discloses the appearance
of nucleation fronts and nucleation events: Waves which enter the test
volume and already carry a temperature disturbance with $T<T_S$ result
in a nucleation front (\eg $x\lesssim 0.025\,l_{\rm ref}$ r.h.s.,
Figs.~\ref{fig:Means}). Waves, \ie turbulence elements, which
interact inside the test volume and only there create $T<T_S$ for a
short time result in nucleation events as the peak like $\langle
J_*\rangle_t$ shows.

Else, the dust quantities are constant in almost the whole test volume
which is in agreement with their time averages. Deviations from these
almost constant values occur only near the volume's boundaries since
here fresh, uncondensed material enters.

{\it Viewing our test volume again as mass element in a convective
  environment which is constantly disturbed by wave propagation, we
  conclude that nucleation will take place everywhere in the mass
  element but likely with very much different efficiency.  Since
  fresh, uncondensed material enters the mass element through open
  boundaries, nucleation can go on here only if the temperature is low
  enough. This does not cause the boundary region to contain the
  largest amount of dust since the dust can also leave the mass
  element if the fluid flow moves outward.}

\begin{figure*}
\begin{center}
\begin{tabular}{cc}
\hspace*{-1.0cm}
\epsfig{file=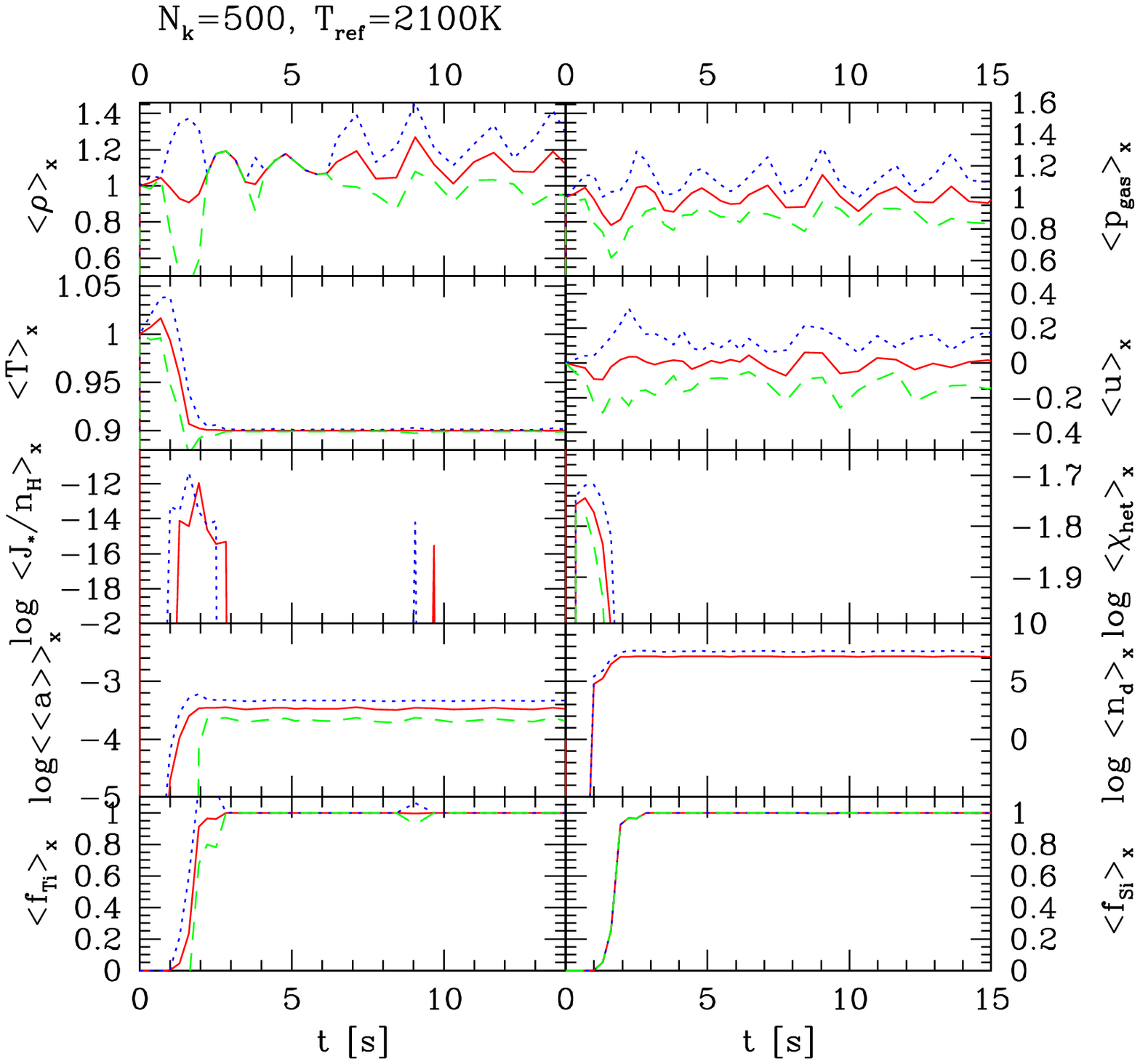, scale=0.55}
&
\epsfig{file=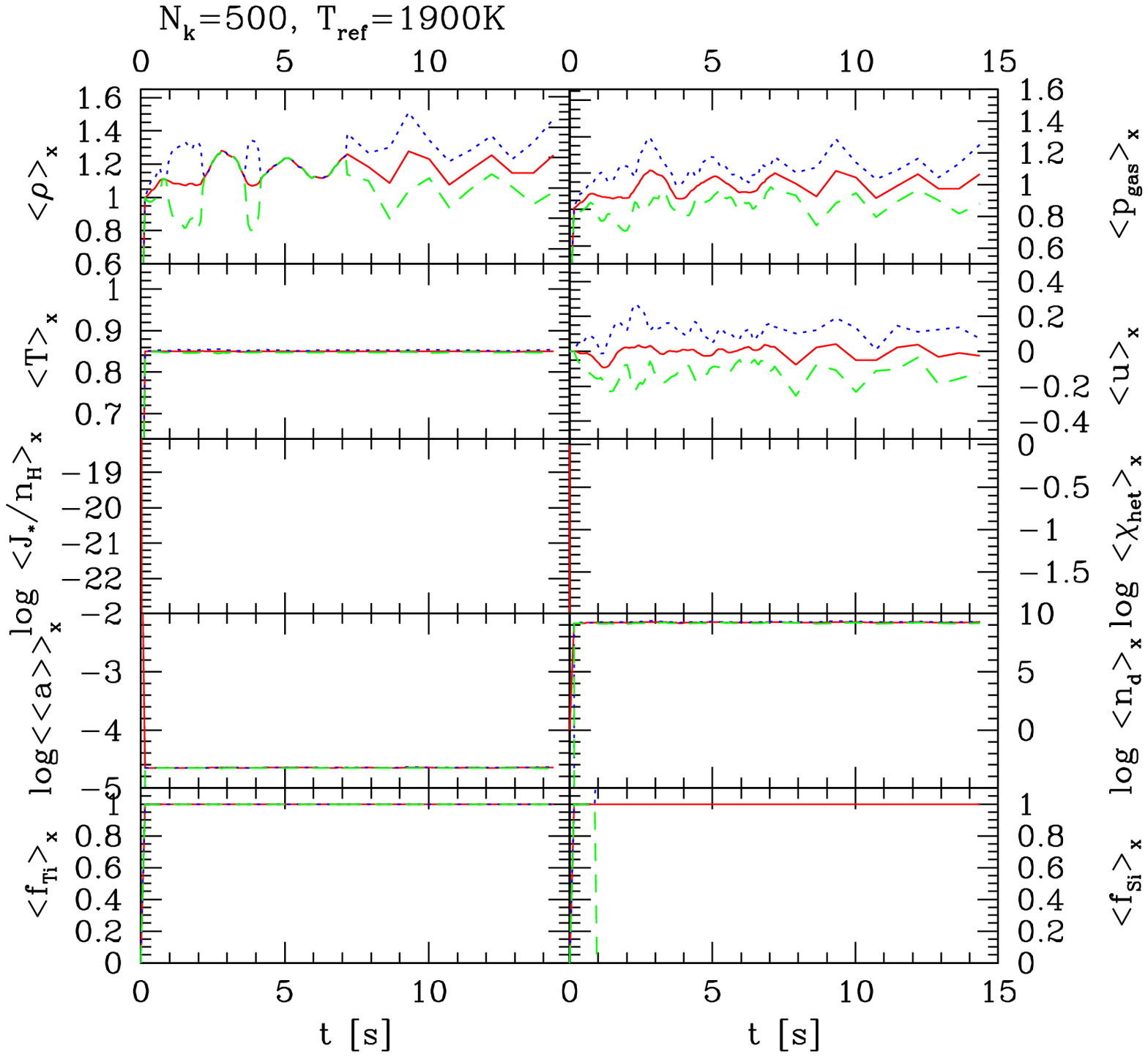, scale=0.55}
\end{tabular}
\caption[]{The space-means $\langle\alpha(t)\rangle_x$ (solid/red) with the apparent standard deviations $\sigma_{\rm N_x-1}^{\alpha}(t)$ (dotted/dashed)  as function of time for $T_{\rm ref}=2100$\, (A, l.h.s.), and $T_{\rm ref}=1900$\, (C, r.h.s.); see Table~\ref{tab:kennzalle}. ($\langle\alpha(t)\rangle_x+\sigma_{\rm N_x-1}^{\alpha}(t)$ - dotted/blue; $\langle\alpha(t)\rangle_x-\sigma_{\rm N_x-1}^{\alpha}(t)$ - dashed/green).}
\label{fig:1Dstoch2100mitStandartdeviation1}
\end{center}
\end{figure*}

\subsubsection{Dependence on the number of modes}\label{subsec:depNk}

Figure~\ref{fig:Nk100_2100K_spacemean} depicts the same calculation
like Figs.~\ref{fig:Nk500_2100K_Ort} --~\ref{fig:Means} but carried out
with different a number of modes, $N_k=100$.  Comparing with the l.h.s.
of Fig.~\ref{fig:Means} ($N_k=500$) shows that the variations in the
hydrodynamic quantities are smaller but the dust quantities reach very
alike mean values independent of $N_k$.

Note, more energy is contained in the small wavenumbers (= large
spatial scales) since the Kolmogoroff spectrum is applied to calculate
the velocity disturbances in the Fourier space.  Consequently, if a
number of chosen modes, $N^1_{\rm k}$, is small, the smallest
wavenumber will contain less energy than the smallest wavenumber for
some larger number of modes $N^2_{\rm k}$, and, from $N^1_{\rm k}<
N^2_{\rm k}$ follows $E(k_1({\rm N^1_k}))< E(k_1({\rm N^2_k}))$.  This
results in the appearance of larger velocity and pressure peaks with
increasing $N_{\rm k}$ according to Eqs.~(\ref{equ:deltauFour}) and
(\ref{equ:deltapFou}).

The study of the long term behaviour of the $T_{\rm ref}=2100\,$K
simulation reveals the occurrence of a long term pattern in $\rho$,
$p$, and $u$ (beat frequency oscillations) with a frequency $\nu_{\rm
  beat}\approx 100$\,s $\approx 1.7$\,min.  Comparing $\rho$, $p$, and
$u$ in Figures~\ref{fig:Nk100_2100K_spacemean} exhibits 6 maxima at
$t\approx 50$s, 150s, 250s, 350s, 450s, 550s. This beat frequency
$\nu_{\rm beat}$ seems independent on the number of modes $N_k$ but
does, however, not establish for an excitation with a very small
number of modes (\eg $N_k=5$, not depicted here) and is smeared for a
very large number of modes due to larger fluctuations around the mean
values ($N_k=500$, Fig.~\ref{fig:Means}).

\begin{figure}
\epsfig{file=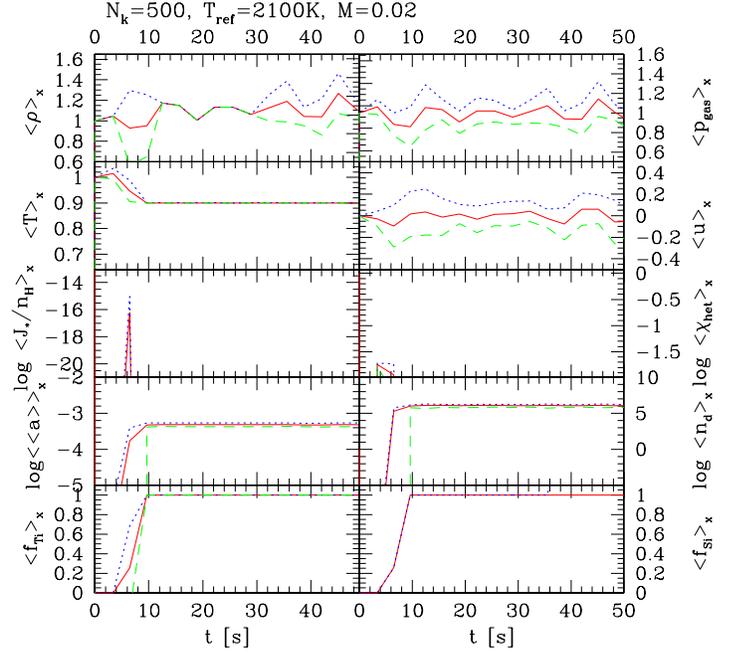, scale=0.54}
\caption[]{Same like Fig.~\ref{fig:1Dstoch2100mitStandartdeviation1} (r.h.s.) but $M=0.02$ ($\nearrow$ entree B Table~\ref{tab:kennzalle})}
\label{fig:1Dstoch2100mitStandartdeviation2}
\end{figure}

\subsubsection{Apparent standard deviation}\label{sssec:standartdev}

Deviations from the most plausible values, the mean values, can be
studied in terms of the apparent standard deviation.
The apparent standard deviation allows to estimate the mean
deviations of characteristic dust quantities due to the turbulent fluid field,

\begin{equation}
\sigma_{N_x-1}^{\alpha}(t):= \sqrt{ \frac{\sum_0^{N_x} (\alpha_i(t))^2 - (\sum_0^{N_x}\alpha_i(t))^2/N_x}{N_x-1}}.
\label{equ:standartabweich}
\end{equation}
Equation~(\ref{equ:standartabweich}) is therefore the mean, quadratic
weighted deviation of the realisations $i$ ($i=0\ldots N_x$) of the
turbulent, dust forming gas flow in space at an instant of time $t$.
Figure~\ref{fig:1Dstoch2100mitStandartdeviation1} depicts the space
means (solid) $\langle\alpha\rangle_x(t)$, and the respective apparent
standard deviations leading to $\langle\alpha\rangle_x(t)+\sigma_{\rm
  N_t-1}^{\alpha}(t)$ (dotted) and
$\langle\alpha\rangle_x(t)-\sigma_{\rm N_t-1}^{\alpha}(t)$ (dashed).
Note that there is no straight forward functional dependence among all
the lower (dashed) and all the upper (dotted) curves, respectively.

\begin{figure}
\begin{center}
\epsfig{file=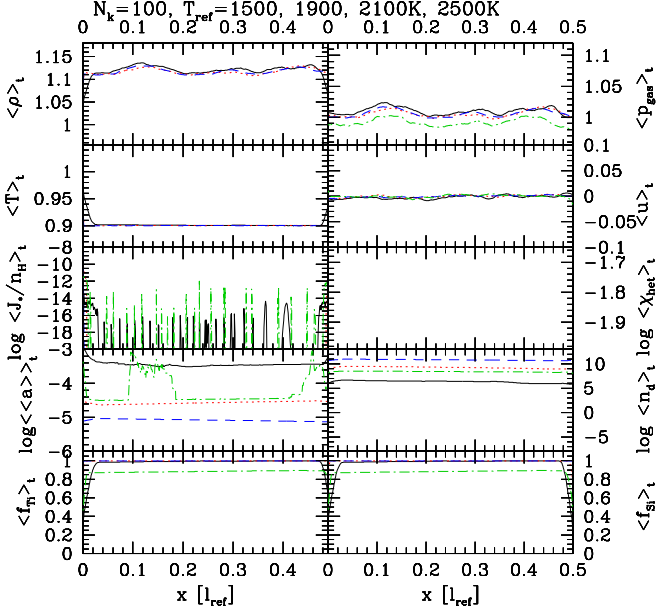, scale=0.5}
\caption[]{Time-means over $\approx 10$min for different temperatures $T_{\rm ref}$: dash-dot (green) - 2500K, solid (black) - 2100K ({\it reference results}), dotted (red) - 1900K, dashed (blue) - 1500K (for details on reference values see Table~\ref{tab:differentTemp})}
\label{fig:1Dstoch}
\end{center}
\end{figure}

Figure~\ref{fig:1Dstoch2100mitStandartdeviation1} shows that the
standard deviation is largest in the period of most active dust
formation, \ie between 0.05\,s and 3\,s for the Mach number case
depicted (compare paragr. {\rm ``Dependence on Mach number''}),
independent on the initial reference temperature. For example, the
minimum and the maximum deviations in density deviate by almost a
factor 3 for the model with $T_{\rm ref}=2100$\,K ($\nearrow$ entree A
in Table~\ref{tab:kennzalle}). The apparent standard deviation of
the velocity field shows that  $\delta u(x,t) = \pm 0.2
v_{\rm ref}$ and less which is subsonic (compare
Figs.~\ref{fig:1Dstoch2100mitStandartdeviation1},~\ref{fig:1Dstoch2100mitStandartdeviation2},~\ref{fig:1Dstoch})
and in agreement with Ludwig et al.~(2002).

The apparent standard deviations indicate that there are no very large
deviations in the 1D dust quantities if the dust complex has reached
it steady state, in contrast to the hydrodynamic quantities. Turbulent
fluctuations will cause the dust formation to set in somewhat earlier
and to occur somewhat more vivid (larger $J_*$). On the contrary,
Figs.~\ref{fig:1Dstoch2100mitStandartdeviation1}
and~\ref{fig:1Dstoch2100mitStandartdeviation2} illustrates that no
dust forms if the fluctuations result in $T>T_S$ (no dashed line for
$J_*$).

\paragraph{Dependence on temperature:}
The standard deviations of the hydrodynamic quantities are not
considerably larger neither in a deeper nor in a shallower turbulence
excited atmospheric layer (r.h.s.
Fig.~\ref{fig:1Dstoch2100mitStandartdeviation1}). In contrast, the
variations in the dust quantities decrease with decreasing $T_{\rm
  ref}$ and increase with increasing $T_{\rm ref}$ (the latter is not
shown here). The nucleation rate decreases by orders of magnitude with
increasing temperature and the standard deviation is considerably
larger.  Consequently, the mean number of dust particle is smaller and
therefore the mean particle size larger. In contrast, nucleation
occurs earlier and more vivid with decreasing temperature. The dust
formation process is complete already during the very first time of
the simulation as \eg depicted on the r.h.s. in
Fig.~\ref{fig:1Dstoch2100mitStandartdeviation1}, therefore $J_*$ and
$\chi_{\rm het}$ are not resolved on the time interval plotted.
Consequently, the number of dust particles is order of magnitudes
higher and the mean particles size has decreased since the material has
to be distributed over a larger grain surface area than if less
particles form.  All of it is in accordance with common expectations (see
Sect.~\ref{subsec:dustwindow}).

\paragraph{Dependence on Mach number:}
Figure~\ref{fig:1Dstoch2100mitStandartdeviation2} shows a simulation
comparable to the r.h.s. of
Fig.~\ref{fig:1Dstoch2100mitStandartdeviation1} but now with $M=0.02$
instead of $M=0.1$ ($\nearrow$ entrees A, B in
Table~\ref{tab:kennzalle}). Consequently, the characteristic time
scale is much longer, namely $t_{\rm ref}\approx 15\,$s instead of
$t_{\rm ref}\approx 3\,$s and a much longer time interval needs to be
depicted in Fig.~\ref{fig:1Dstoch2100mitStandartdeviation2} compared
to Fig.~\ref{fig:1Dstoch2100mitStandartdeviation1} in order to observe
the inset of the dust formation. 

We observe that the variation of the hydrodynamic quantities does not
change remarkable compared to higher Mach number cases. It only
appears on a much longer time scale.  However, the superimposed waves
need about 3 times longer to initiate the first dust formation.  The
nucleation is somewhat less efficient resulting in a slightly lower
number of dust particles which are, hence, slightly larger.  Also the
growth process is less efficient compared to the M=0.1--case

Although the dust complex acts on its own, chemical time scales, it
needs much longer time to reach a steady state situation if the initial
Mach number is small (see also Fig.~\ref{fig:1DTauKappa} in
Sect.~\ref{subsec:variability}).

\subsection{The dust formation window}\label{subsec:dustwindow}

Stochastic fluctuation can drive a reactive gas flow into the dust
formation window, \ie the thermodynamic regime where the gas - solid
(or liquid) phase transition is possible and most efficient (see \eg
Sedlmayr 1997)\nocite{sed97}.

Depending on the thermodynamic (TD) situation, three regimes (compare
Fig.~\ref{fig:regimes}) appear to be present in a turbulent
atmosphere: The {\it deterministic} (or {\it subcritic}) contains
those TD states where dust formation occurs without the need of an
(\eg hydrodynamic) ignition, i.e. the local temperature is already
smaller than the nucleation threshold temperature.

The {\it stochastic} regime contains those TD states for which the
dust is possible if some realistic ignition mechanism can cause
$T<T_{\rm s}$.  This regime contains the critical range where a
transition from $T>T_{\rm s}$ to $T<T_{\rm s}$ is possible. The size
of the stochastic regime depends on the turbulent energy.

The third regime can be called {\it impossible} since no dust
formation would be possible here.

\paragraph{Temperature dependence:}
We have investigated the transition deterministic -- stochastic regime
by studying the temperature dependence in our stochastic 1D
simulations.  The time mean values (Fig.~\ref{fig:1Dstoch}) of the
dimensionless hydrodynamic variables are very much alike
($\langle\rho\rangle_t$, $\langle p\rangle_t$, $\langle T\rangle_t$,
$\langle u\rangle_t$, for reference values see
Table~\ref{tab:differentTemp}) but the dust quantities deviate
considerably between these two extreme regimes\footnote{The difference of
  $\langle p\rangle_t$ in the case of $T=2500$K is correct since here
  $T_{\rm RE}/T_{\rm ref}$ had to be considerably smaller in order to
  allow the system to enter the dust formation window (compare
  Table~\ref{tab:differentTemp}).}.

Figure~\ref{fig:1Dstoch} depicts four cases of which $T_{\rm
  ref}=1900, 1500$K (dotted, dashed) fall into the deterministic
regime in which the dust formation process is complete after a very
short time on the whole test volume ($f_{\rm Ti}=f_{\rm Si}=1$)
without any external excitation necessary. The $T_{\rm ref}=2500,
2100$K (dash-dot, solid) fall into the stochastic regime where
turbulence initiates the dust formation process by causing the very
first nucleation event to occur (compare also Sect.~\ref{sec:Intro}).
In $T_{\rm ref}=2100$K-case the dust formation process is still
completed after a very short time ($t\approx 0.07$s) in the whole test
volume while for $T_{\rm ref}=2500$K the first efficient nucleation
event occurs only after about $65$\,s\,$\approx 1$min. The dust
formation is not complete ($f_{\rm Ti}, f_{\rm Si}<1$) in this
comparably hot case and even much more refined in time and space: The
time mean of the mean particle size,$\langle\langle a\rangle
\rangle_t$, varies by $\approx 1$ order of magnitude (4th panel,
l.h.s, Fig.~\ref{fig:1Dstoch}). Therefore, $T_{\rm ref}=2500\,$K falls
in the very end of the stochastic regime being close to impossible.

\begin{figure*}
\vspace*{-0.2cm}
\begin{tabular}{ll}
\hspace*{2cm}{\large $\log\,n_{\rm d}$ [cm$^{-3}$]} and vorticity $(\nabla\times \boldsymbol{v})$ [s$^{-1}$] & \hspace*{2cm}{\large $\log \langle a \rangle$ [cm]}\\[0.7cm]
\hspace*{-0.5cm}
\epsfig{file=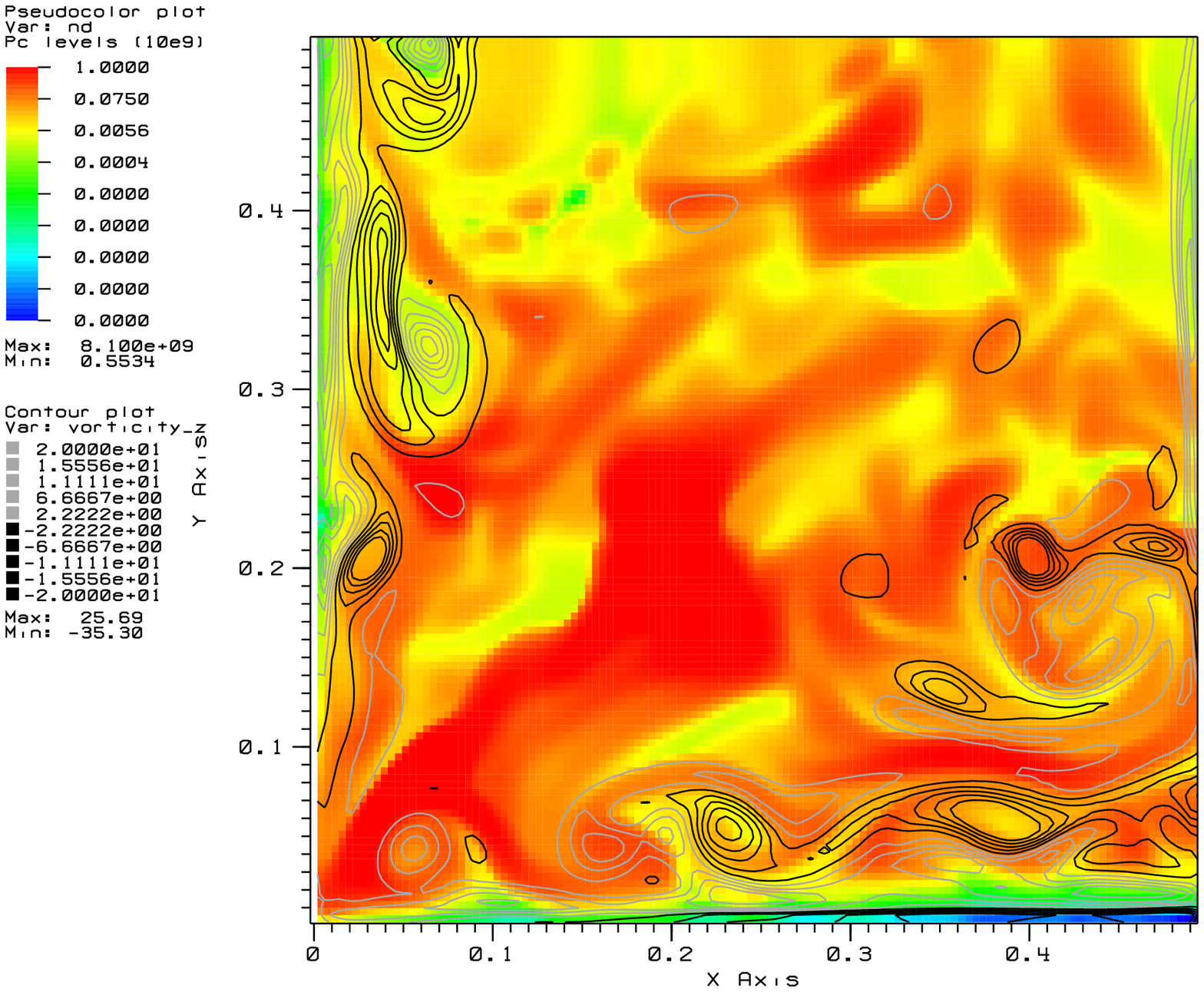, scale=0.45}
&
\hspace*{-0.5cm}
\epsfig{file=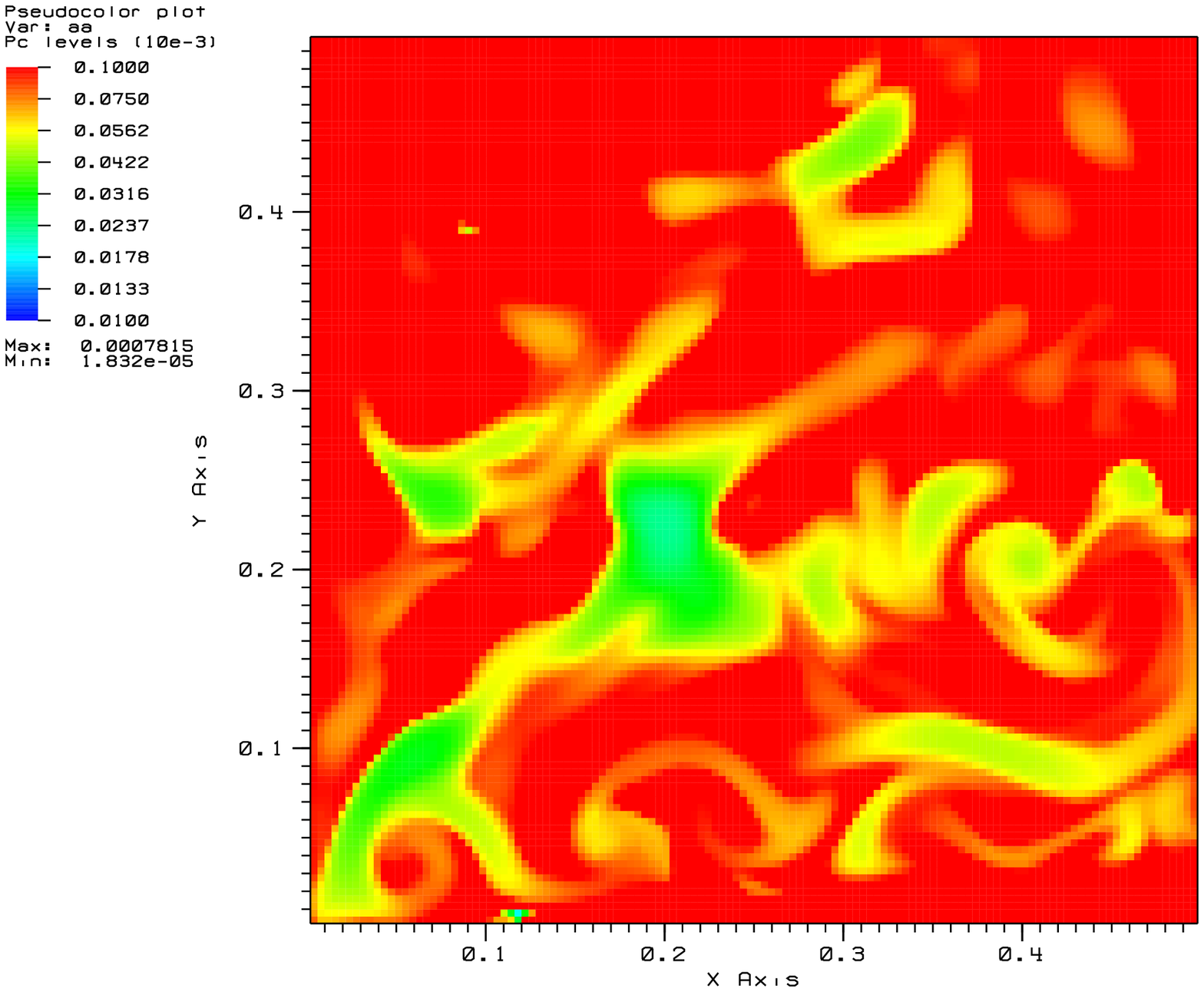, scale=0.45}\\[3.2cm]
\hspace*{2cm}{\large $\log\,n_{\rm d}$ [cm$^{-3}$]} & \hspace*{2cm}{\large $\log \langle a \rangle$ [cm]}\\[0.7cm]
\hspace*{-0.5cm}
\epsfig{file=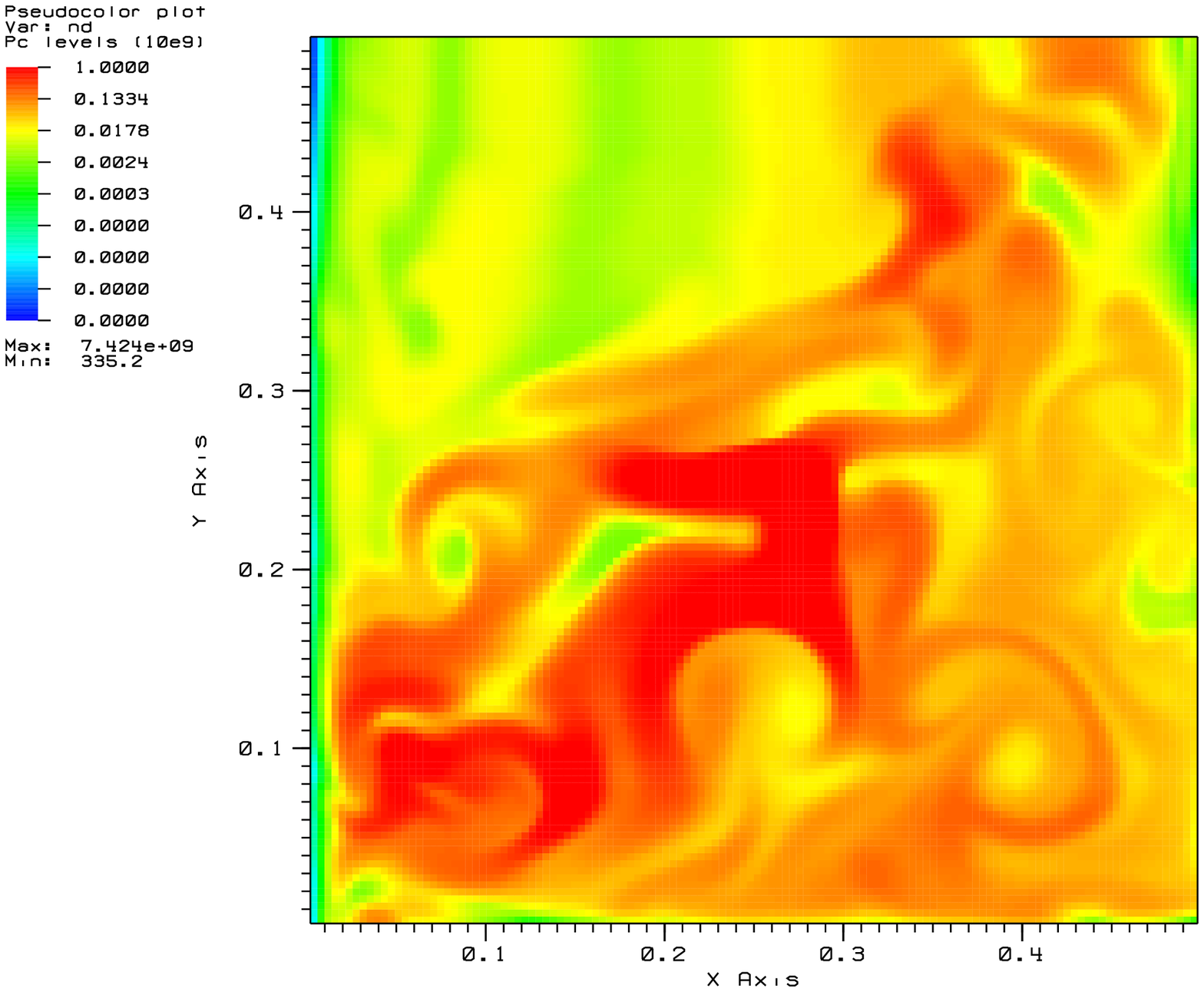, scale=0.45}
&
\hspace*{-0.5cm}
\epsfig{file=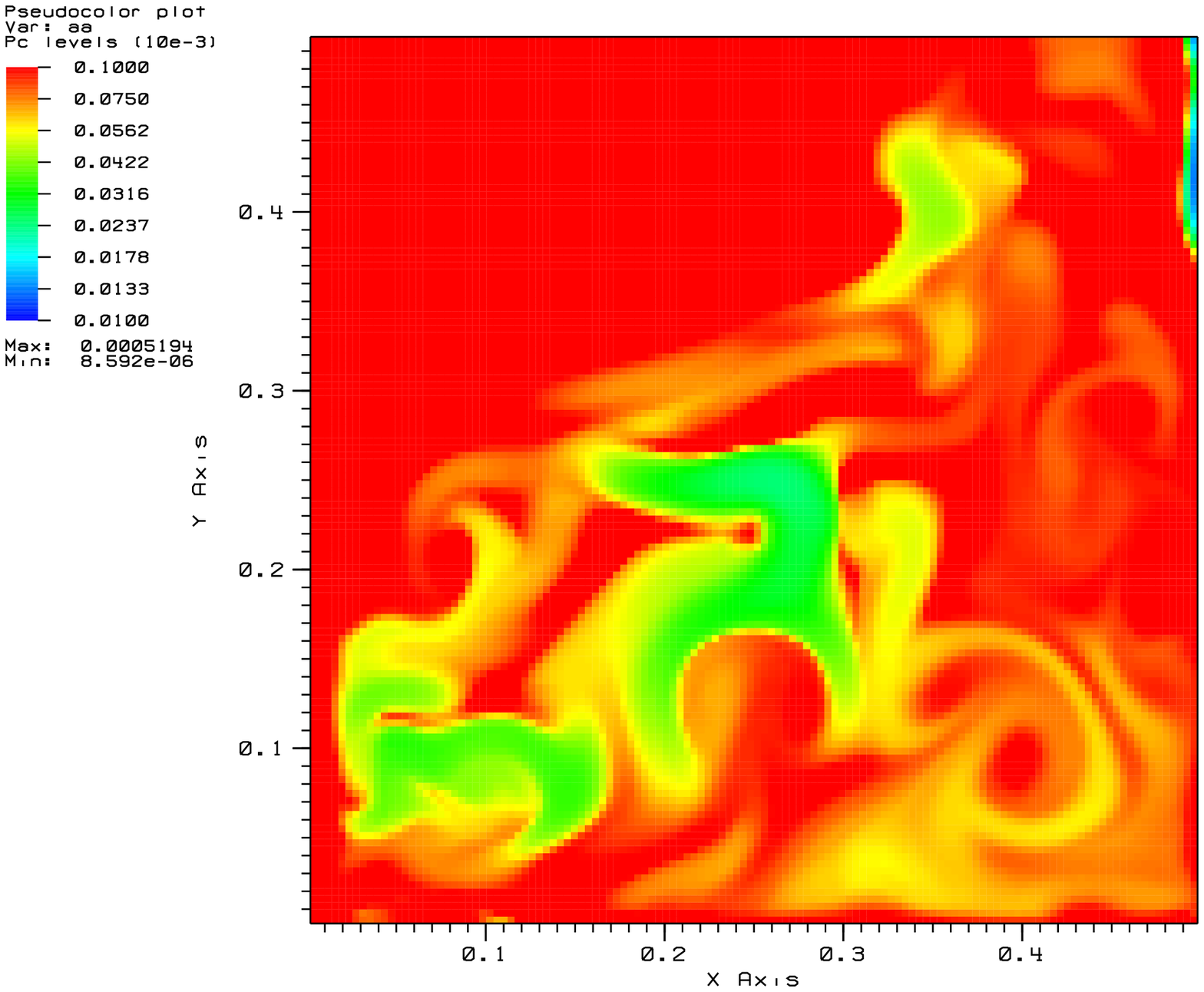, scale=0.45}\\[3.2cm]
\hspace*{2cm}{\large $\log\,n_{\rm d}$ [cm$^{-3}$]} & \hspace*{2cm}{\large $\log \langle a \rangle$ [cm]}\\[0.7cm]
\hspace*{-0.5cm}
\epsfig{file=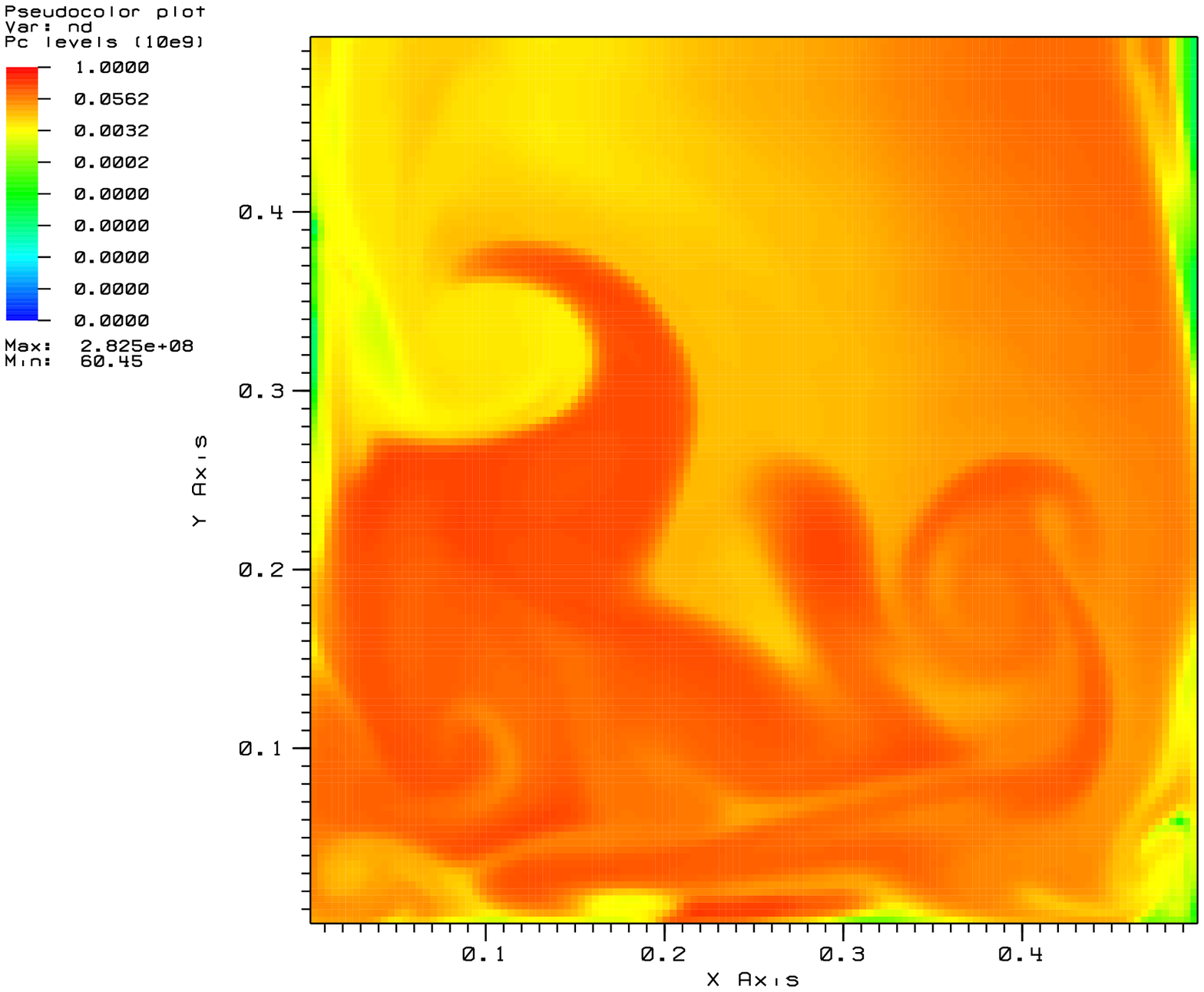, scale=0.45}
&
\hspace*{-0.5cm}
\epsfig{file=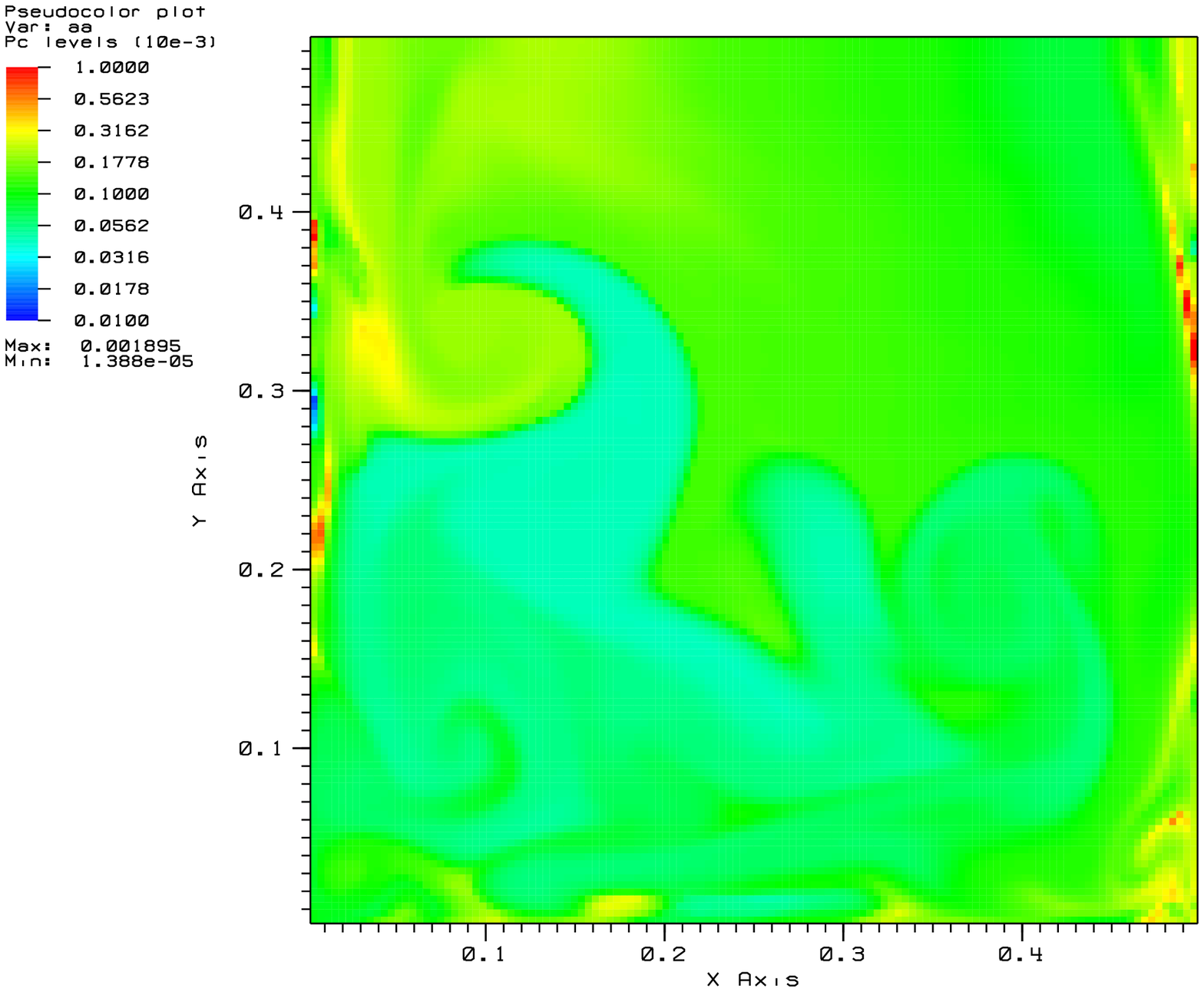, scale=0.45}\\[3.2cm]
\end{tabular}
\caption[]{2D simulations with $T_{\rm ref}=2100$\,K, $M=1$, 
 $N_{\rm k}=500 (\,\,\nearrow\,\,$entree A$^{\prime}$ Table~\ref{tab:kennzalle}) for three instants of  time during 
 the period of active dust formation (top: 0.8s, middle: 1.7s, bottom: 8s; $N_{\rm x}\times N_{\rm y}= 128\times 128 = 500\,\mbox{m} \times 500\,$m, $\boldsymbol{g} = \{0, -g, 0\}$)}
 {\bf Left}: number of dust particle $\log\,$n$_{\rm d}\,$ [cm$^{-3}$] and vorticity $(\nabla\times \boldsymbol{v}$) [s$^{-1}$] only for $t=0.8$s; 
 {\bf Right}: mean particle radius $\log\,\langle a\rangle$\, [cm].
\label{fig:2D}
\end{figure*}

\begin{table}[h]
\begin{center}
\caption[]{$\rho_{\rm ref}=3.16\,10^{-4}$ g\,cm$^{-3}$, $l_{\rm ref}=0.5\,10^{-5}$cm}
\label{tab:differentTemp}
\begin{tabular}{c|c|c|c|c}
\hline
$T_{\rm ref}$ & $p_{\rm ref}$    & $u_{\rm ref}$  & $t_{\rm ref}$ & $T_{\rm RE}$\\
      $[$K$]$ & [dyn\,cm$^{-2}$] & [cm\,s$^{-1}$] & [s]           & $[$K$]$\\
\hline
2500 & $2.84\,10^{7}$ & $3.54\,10^{4}$ & 2.821 & 1750 $(=0.7\,T_{\rm ref})$\\
2100 & $2.38\,10^{7}$ & $3.25\,10^{4}$ & 3.078 & 1890 $(=0.9\,T_{\rm ref})$\\
1900 & $2.16\,10^{7}$ & $3.09\,10^{4}$ & 3.236 & 1710 $(=0.9\,T_{\rm ref})$\\
1500 & $1.70\,10^{7}$ & $2.74\,10^{4}$ & 3.642 & 1350 $(=0.9\,T_{\rm ref})$\\
\hline
\end{tabular}
\end{center}
\end{table}

The temperature sequence depicted in Fig.~\ref{fig:1Dstoch} displays
the transition from the deterministic into the stochastic regime: the
dust formation is most efficient at the smallest temperature
considered (1500K) resulting in the most dust particles (3rd panel,
r.h.s., Fig.~\ref{fig:1Dstoch}) and therefore in the smallest grain
size. With increasing temperature, less particles are formed which can
accumulate considerably more material and grow therefore to largest
sizes. The hottest case considered seem not to fit into this picture
because \eg  $\langle\langle a\rangle \rangle_{t, 2500K} >
\langle\langle a\rangle \rangle_{t, 1500K}$ but $\langle\langle
a\rangle \rangle_{t, 2500K} \approx \langle\langle a\rangle
\rangle_{t, 1900K}$. Comparing the radiative equilibrium temperature
of our test calculations (Table~\ref{tab:differentTemp}) indicates
that $T_{\rm RE, 2500K}\approx T_{\rm RE, 1900K}$. Apart from the fact
that considerably more time is needed to form and grow the first,
initiating seed particles the higher the temperature is, the dust will
drive the system locally towards this low radiative equilibrium
temperature providing thereby TD conditions comparable to the
$T=1900$K-case.  The resultant dust quantities need therefore be
comparable if $T=T_{\rm RE}$, \ie if the gas has reached the same
isothermal state.


\begin{figure*}
\begin{center}
\begin{tabular}{cc}
\hspace*{-1.0cm}
\epsfig{file=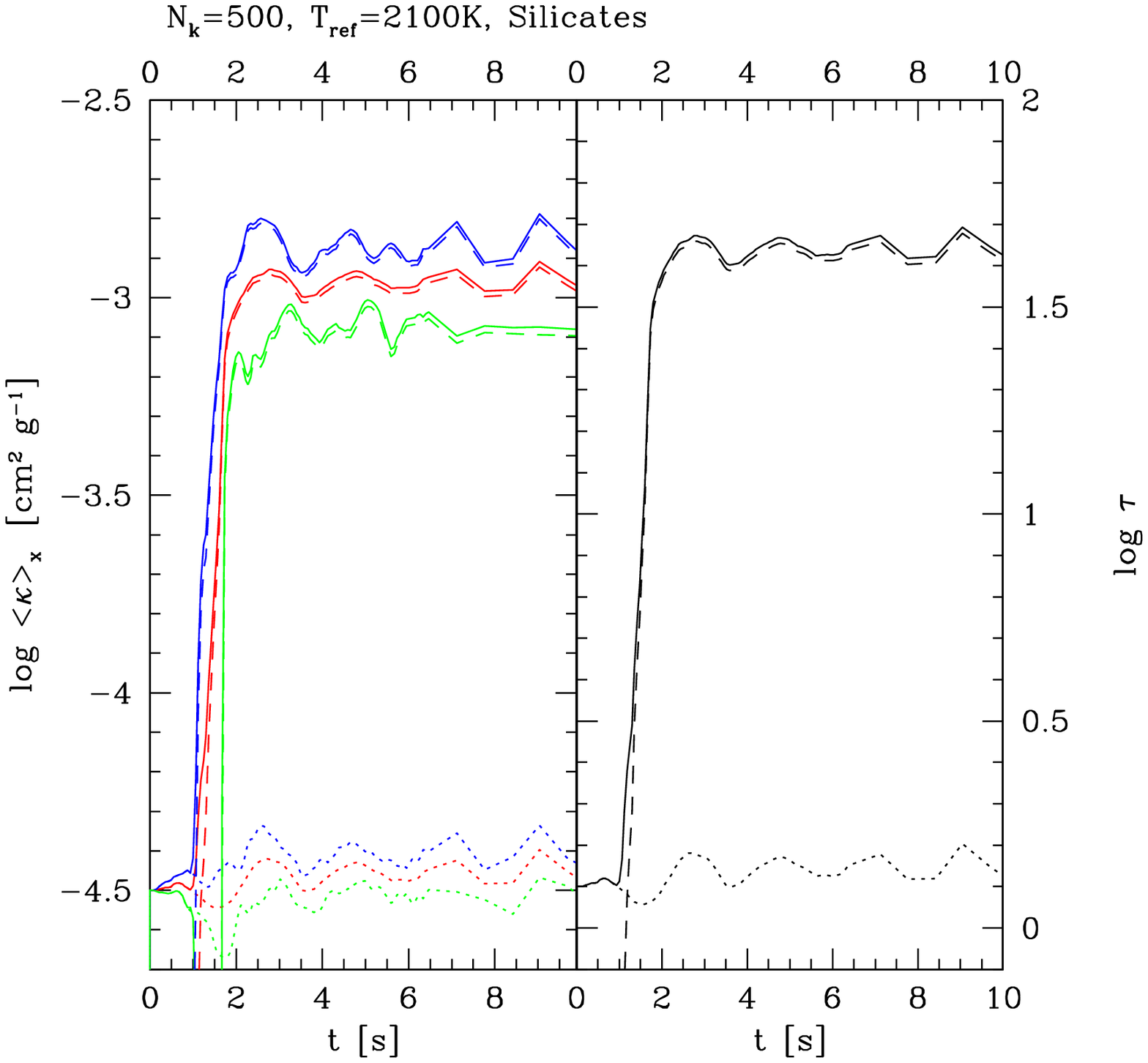, scale=0.5}
&
\epsfig{file=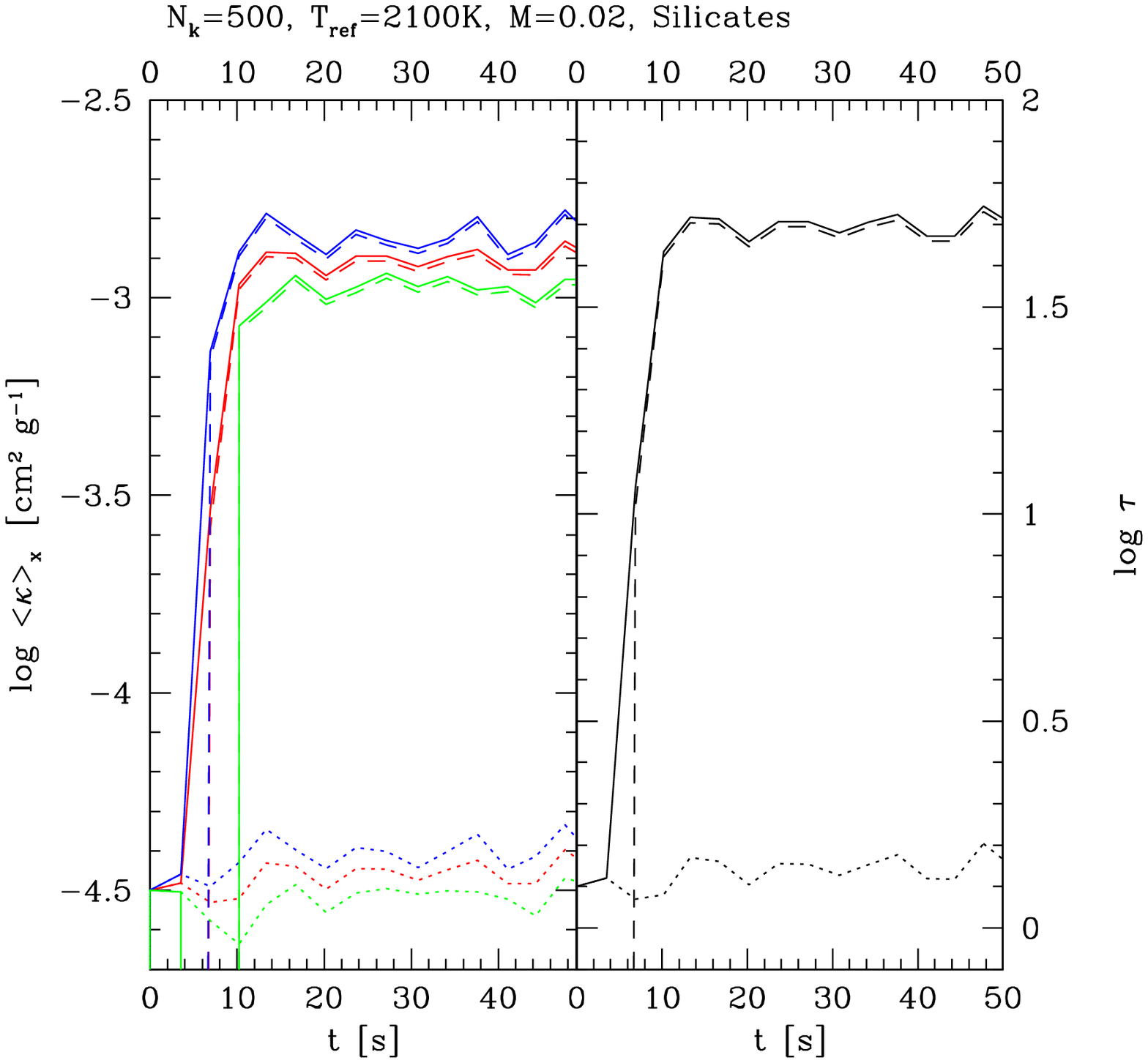, scale=0.5}
\end{tabular}
\caption[]{Space mean and standard deviations (solid) of the total grey
absorption coefficient $\kappa=\kappa_{\rm gas}+\kappa_{\rm dust}$ and
the total grey optical depth $\tau$ for the simulations depicted in
Fig.~\ref{fig:1Dstoch2100mitStandartdeviation1}. The dust (dashed)
and the gas (dotted) contributions are shown for simulations with
$T_{\rm ref}=2100$\,K with $M=0.1$ (l.h.s.; entree A
Table~\ref{tab:kennzalle}) and with $M=0.02$ (r.h.s.; entree B
Table~\ref{tab:kennzalle}).}
\label{fig:1DTauKappa}
\end{center}
\end{figure*}

\subsection{2D results}\label{subsec:2D}

So far, only 1D results have been presented in this paper, which
provide a good possibility to study the most important ongoing
physical and chemical processes and their interactions. In 1D,
however, each wave crosses the whole test volume and will therewith
influence the local thermodynamic conditions everywhere inside the
volume. In 2D, the influence of waves leads to much more complicated
patterns since \eg a non-zero rotation of the fluid field can
develop.  The expected consequence is a much more heterogeneous
distribution of dust than in any 1D situation as it is illustrated in
the following.

A 2D model calculation with $T_{\rm ref}=2100$\,K, $M=1$, $N_{\rm
k}$=500 ($\,\,\nearrow\,\,$ entree A$^{\prime}$ in
Table~\ref{tab:kennzalle})\footnote{Low Mach number simulations
of driven turbulence in 2D are not yet possible with the present
code. Botta\etal (2003)\nocite{bkll2003} have shown that unbalanced
truncation errors can lead to considerable instabilities in the
complete, time-dependent equation of motion in a quasi-static
situation and suggest a balanced discretisation scheme. We will tackle
this problem in a forthcoming paper and use our present 2D results for
M=1 only to illustrate the stronger influence of the hydrodynamic
processes on the evolving dust structures to be expected in
multi-dimensional simulations compared to 1D.} was performed on a
spatial grid of $N_{\rm x}\times N_{\rm y} = 128\times 128$ cells
corresponding to a box of 500m $\times$ 500m.  The smallest eddies
have a size of $\lambda^{\rm 2D}_{\rm min}=5\,$m, the largest are of
the size of the test volume.  The gravity acts in negative
$y$-direction, \ie $\boldsymbol{g} = \{0, -g, 0\}$. The initially
homogeneous and dust free fluid is constantly disturbed by
superimposed waves entering from the left, the right, and the bottom
side. Thereby, a gas element is modelled which is continuously
disturbed by waves originating from the surrounding convectively
instable atmospheric fluid. The top side is kept open simulating the
open upper boundary of a test volume in the substellar atmosphere.
  
Figure~\ref{fig:2D} shows three instants of time during the phase of
vivid dust formation and demonstrates the appearance of large and
small dusty scale structures evolving with time. Both, the number of
dust particles $n_{\rm d}$ (l.h.s.) and the mean particle size
$\langle a\rangle$ (r.h.s.), are plotted on a logarithmic scale with
$n_{\rm d} = 1\,\ldots\,10^9$ cm$^{-3}$ and $\langle a\rangle =
10^{-5.5} \,\ldots\,10^{-3.5}$. The very inhomogeneous appearance of
the dust complex is a result of nucleation fronts and nucleation
events comparable to our 1D results. The nucleation is now triggered
by the interaction of eddies coming from different directions. Large
amounts of dust are formed and appear to be present in lane-like
structures (large $\log\,$n$_{\rm d}$; dark areas). The lanes are
shaped by the constantly inward travelling waves. Our simulations show
that some of the small-scale structure merge thereby supporting the
formation of lanes and later on even larger structures. The formation
of such large structures is not caused by the establishment of a
pressure gradient to counterbalance the gravity.  Since the whole
test volume is only of the size of $H_{\rm P}/20$ the resulting
pressure gradient is negligibly small. Hence, the large-scale
structures result from the interaction of dust formation and
turbulence.
  
Furthermore, dust is also present in curl-like structure which
indicates the formation of vortices.  As the time proceeds in our 2D
simulation, vortices develop orthogonal to the velocity field which
show a higher vorticity ($\nabla \times \boldsymbol{v}(\boldsymbol{x},
t)$) than the majority of the background fluid field.  For
illustration, the maximum and the minimum vorticity between $\approx
-20$\,s$^{-1}$ and $\approx 20$\,s$^{-1}$ has been superimposed as a
contour plot (grey/black) on top of the false colour plot of the
number of dust particles for $t=0.8$\,s in Fig.~\ref{fig:2D} (l.h.s.,
top).  This shows that the vortices with high vorticity preferentially
occur in dust free regions or regions with only little amounts of dust
present. The motion of the vortices can transport the dust particles
into region with still condensible material available, and seem
thereby to cause larger and larger dusty areas to form.

The 2D hydrodynamic behaviour is comparable to our 1D results in the
sense that the stochastically created waves enter the test volume,
interact, run through the test volume, and eventually leave it at the
top side after they have initiated the dust formation process.  Large
changes in the dust quantities during short time intervals occur
locally during the time period of the first nucleation and the
re-initiation of nucleation by radiative cooling. A spatially
inhomogeneous dust distribution results which is only slightly shifted
back and forth by the inward moving superimposed waves. In contrast,
considerable variations in $\rho$, $p$, and in the velocity components
$(u, v)$ occur in time and space. Furthermore, a new qualitative
behaviour compared to the 1D simulation appears. Hydrodynamic
advection gathers the dust in larger and larger structures while the
formation of the dust has been initiated in the small-scale regime of
the simulation. The reference time of our present 2D simulation is
comparably high (see Table~\ref{tab:kennzalle}). We have, however,
done 1D test calculations
(Figs.~\ref{fig:1Dstoch2100mitStandartdeviation1},~\ref{fig:1Dstoch2100mitStandartdeviation2},~\ref{fig:1DTauKappa})
which indicate that in the low Mach number case the global dust
formation time scale (\ie the time when the dust formation reaches its
steady state) will merely increase (for discussion see
Sect.~\ref{subsec:upscale}).

\section{Discussion}\label{sec:disc}

\subsection{Variability}\label{subsec:variability}
Observational evidence has been provided (Bailer-Jones \& Mund 1999,
2001a,b; Mart\'{i}n et al.~2001; Gelino et al.~2002; Clarke et al.
2002)\nocite{bm99,bm2001a,bm2001b,mzl2001, gm2002, clark2002} that
Brown L-Dwarfs are non-periodically photometric variable at a low
level of $1-2\%$ (Clarke 2002)\nocite{clark2002} and sometimes even
larger.  Nakajima et al~(2000) and Kirkpatrick et
al.~(2001)\nocite{ntmim2000, kdmrglb2001} reported on spectroscopic
variability. Appealing explanations are the appearance of magnetic
spots or the formation of dust clouds.  Mohanty et al.~(2002) and
Gelino et al.~(2002)\nocite{mbsag2002, gmhal2002} have argued that
ultra cool dwarfs are unlikely to support magnetic spots. There
is, however, evidence for magnetic activity in L dwarfs by a rapid
declining, strong H$\alpha$ emission. In contrast, three objects are
observed with a persistent, strong H$\alpha$ emission
(Liebert\etal~2003)\nocite{}. A straight forward, consistent
explanation is not at hand yet and will probably be theoretically very
demanding. We, therefore, follow in this paper the hypothesis of the
formation of dust clouds in a convectively influenced turbulent
environment as explanation of non-periodic variability

Ludwig et al.~(2002)\nocite{lah2002} argue
based on their hydrodynamic 3D simulation that M-dwarfs (and even more
Brown Dwarfs) show only very little temporal and horizontal
fluctuations in their atmospheres. 
Dust strongly interacts with the thermo- and hydrodynamics due to
radiative transfer effects, gas phase depletion and on macroscopic
scales due to drift. It seems therefore likely to expect a support of
initially small inhomogeneouties by these processes. Woitke (2001) has
carried out 2D radiative transfer calculations for a inhomogeneous
density distribution which support this idea. Since the radiation is
blocked by condensing dust clouds of sufficient optical depth, the
radiation is forced to escape mainly through the remaining wholes,
thereby enhancing and preventing the dust formation, respectively.

One may speculate that the consideration of dust formation in 3D
simulation may even cause these models to deviate considerably from
the simple MLT models in the case of brown dwarfs due to the time
dependence of the dust formation process and the corresponding
feedback on the space and time evolution.

\paragraph{Optical depth:}

Figure~\ref{fig:1DTauKappa} (left panels) depicts the space mean total
opacity (solid lines) and the mean opacity of the dust (dashed lines)
and the gas (dotted lines) for the 1D test calculation investigated in
Fig.~\ref{fig:1Dstoch}. Upper curves indicate
$\langle\kappa\rangle_x(t)+\sigma_{\rm N_t-1}^{\kappa}(t)$, the lower
depict $\langle\kappa\rangle_x(t)-\sigma_{\rm N_t-1}^{\kappa}(t)$.
The Rosseland mean {\it dust} opacities for astronomical silicates
($\kappa_{\rm dust}= 0.75\,\rho L_3 1.74\,T^{1.12}$, Paper~I; l.h.s.)
and a typical Rosseland {\it gas} mean of $\kappa_{\rm
  gas}=0.1$\,g/cm$^3$ have been adopted.  The Rosseland gas mean
opacity was chosen typical for a hot, inner layer of a brown dwarf
atmosphere. Figure~\ref{fig:1DTauKappa} shows that the dust and the
gas opacities differ by about 1.5 orders of magnitude. The dust
opacities varies by about 0.5 magnitudes, the gas opacity by only
about 0.2 order of magnitudes. this is independent of the
characteristic time scale of the system, \ie the large-scale Mach
number of the initial configuration (compare l.h.s. and r.h.s. in
Fig.~\ref{fig:1DTauKappa}).

Figure~\ref{fig:1DTauKappa} further depicts in the right panels the
time evolution of the space mean of the optical depth which a mass
element of the size of our test volume (500\,m) would have. The
optical depth increases by a factor of 10 when the dust forms in
accordance with the opacity increase. There is, however, a delay
between the inset of dust formation and the time when the maximum
optical depth is reached because the dust formation sets in earlier
than a cloud becomes optically thick. This delay is sensitive to the
characteristic time scale of the system as a comparison of the l.h.s.
and r.h.s. in Fig.~\ref{fig:1DTauKappa} demonstrate.

The gas/dust mixture will only be optically thick in the case of
non-transparent dust particles (\eg astronomical silicates) which may
depend on the wavelength considered. Glassy grains may efficiently
absorb in the far IR ($\lambda\ga 10\mu$m) and one might, depending on
the observed wavelength, detect typical dust features and possibly
even the rapid formation events which, however, may need to
evolve on macroscopic scales in order to be observable.

Turbulent fluctuation seem not to produce considerable variations in
view of the space mean of the optical depth in the long term (here
$t>4$\,s) if astronomical silicates are assumed as typical dust
opacity carriers.

\subsection{Towards the observable regime}\label{subsec:upscale}

The time-dependent simulations performed so far might suggest an
up-scaling of the achieved results in order to estimate e.g. possible
cloud sizes and variability time scales on an observational
level. This idea is not as straight forward as it might seem because
of the differences between chemical and hydrodynamical time
scales. Dust formation is a local process and its (local) time scale
is the same independent of the hydrodynamical regime but the
hydrodynamic time scale changes depending on \eg the characteristic
length of the regime considered. A simple and correct up-scaling
according to the principle of similarity would require the
characteristic numbers to remain the same which is not the case (see
also Helling 2003, Lingnau 2004\nocite{hel2003, li2004}), for instance
due to the changing characteristic hydrodynamic time scales. One might
further imagine to upscale the results by a periodic continuation,
i.e. each small-scale simulation being one pixel of the large-scale
picture.  In any case, the feedback of the large scale
(e.g. convective) motions on the small scales is neglected and visa
versa. Important effects like chemical mixing and kinetic energy input
into the turbulent fluid field are than missing.  Consequently, the
non-linear coupling between dust formation and hydrodynamics makes it
impossible to simply scale up detailed small-scale results into an
regime accessible by observations. The complete large-scale simulation
needs to be awaited.

The aim of the investigations of the small-scale regimes performed so
far was indeed to provide information and understanding for building a
large-scale model of a brown dwarf atmosphere.  From these results, we
suggest the following necessary criteria for a sub-grid model (also
called closure term or closure approximation) of a turbulent, dust
forming system:
\begin{itemize}

\item[{\bf a)}] A sub-grid model must describe the transition
  deterministic $\rightarrow$ stochastic dust formation, depending on
  the turbulent energy as \eg measured by the large-scale variations of
  the local velocity field.

\item[{\bf b)}] The dust formation process (nucleation + growth) is
  restricted to a short time interval (of the order of a few seconds),
  which is usually much smaller than the large-scale hydrodynamic time
  scale. This involves that:
  \begin{itemize}

  \item[$\bullet$] The nucleation occurs locally and event-like
    in very narrow time slots.

  \item[$\bullet$] The growth process continues as long as condensible
    material is available and thermal stability of the dust is assured.
    
  \item[$\bullet$] The condensation process finally freezes in and the
    inhomogeneous dust properties are preserved. 

  \end{itemize}
  
\item[{\bf c)}] The dust formation process should be accompanied by a
  fast transition from an approximately adiabatic to an approximately
  isothermal behaviour of the dust/gas mixture. This transition can be
  expected to affect the convective stability in substellar atmospheres.

\end{itemize}

\subsection{Comparison with classical model atmospheres}\label{subsec:classatm}

\begin{figure}
\begin{center}
\epsfig{file=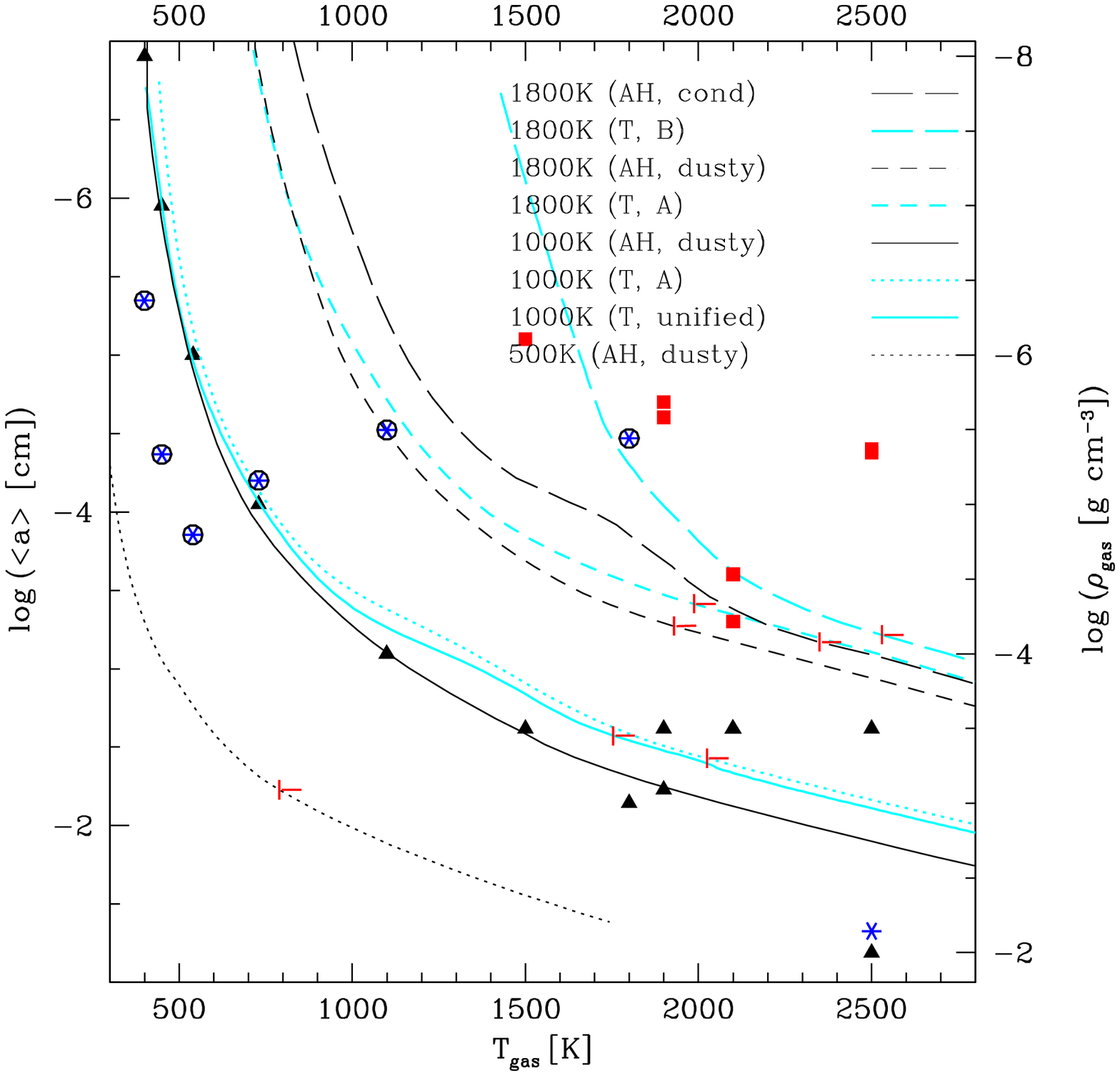, scale=0.45}
\caption[]{The deterministic and the stochastic dust formation regime
  in comparison to typical substellar model atmospheres (solar
  metalicity, $\log g=5$, AH (black) = Allard\etal 2001, T (grey/cyan) = Tsuji~2002).\\
  (open circles (black) - dust formation without ignition; asterisk
  (blue) - dust growth on initially present seed particles
  (1cm$^{-3}$); squares (red) - dust formation under turbulent
  conditions; triangles (black) - $(T_{\rm ref},\rho)$ pairs
  investigated; hooks - border of convective zone)}
\label{fig:0D}
\end{center}
\end{figure}

Figure~\ref{fig:0D} revisits the idea of the dust formation window but
now with view on classical brown dwarf atmosphere calculations. 

The following calculations are performed for various $(T, \rho)$ pairs
in order to demonstrate in which atmospheric regions dust formation
will simply take place (deterministic regime) and in which regions
dust formation needs to be initiated (stochastic regime; compare
Sect.~\ref{subsec:dustwindow}). For eye guidance, the $(T, \rho)$
pairs are indicated by small black triangles along with presently used
substellar model atmospheres in the literature (Allard\etal 2001 --
black lines, Tsuji~2002 -- grey/cyan
lines)\nocite{tsu2002b,ahats2001}. One may notice that the warm models
of the same stellar parameter can differ by about 500\,K for a given
density in the inner atmospheric region which is convectively unstable
in both cases.  Small hooks (red) indicate the upper Schwarzschild
boundary (SB) for convection if available, \ie $v_{\rm MLT}=0$ for
$T<T_{\rm SB}$. The test is simple: Two calculations are carried our
for each $(T, \rho)$ pair where in the first test a small number of
seed particles is prescribed ($\rho L_0=1$cm$^{-3}$; open black
circles), and in the second test no seed particles are prescribed
(blue asterisks). The results should be almost identical inside the
deterministic regime but no dust should form in the stochastic regime.

One observes that there is no difference in grain size in the cool
outer atmospheric region whether seed particles are initially present
or not. The initially present little number of seed particles is
overran by a very efficient nucleation which is characteristic for the
deterministic regime. Note that the dust grains will immediately start
to move inwards on macroscopic scales of atmospheric extension due to
the large gravity of brown dwarfs. However, depending on the TD
conditions, the nucleation efficiency varies in the atmosphere. The
more efficient nucleation takes place, the more seed particles are
formed resulting in smaller mean sizes because the gaseous material
has to be distributed over more particles than in case of less
efficient nucleation. Small mean particle sizes inside the
deterministic regime are therefore a sign for efficient dust
nucleation.

For $T>1800$\,K, the stochastic regime is entered. Here, dust is only
present if by some mean a certain number of seeds is present (blue
asterisks) or a ignition mechanism (\eg turbulence, radiative cooling
by gas) provides the appropriate TD conditions (red squares). Note
that no open circles appears at this site of Fig.~\ref{fig:0D} since
the TD conditions are now inappropriate for a gas - solid phase
transition except an ignition mechanism is present.

\section{Conclusions}\label{sec:concl}

We have studied the onset of the dust formation process in a
turbulent fluid field, typical for the dense and initially
dust-hostile regions in substellar atmospheres. The main scenario is a
convectively ascending fluid element in a brown dwarf atmosphere, which
is excited by turbulent motions and just reaches sufficiently low
temperatures for condensation, but other applications are also
conceivable. The dust formation in a turbulent gas is found to be
strongly influenced by the existence of a nucleation threshold
temperature $T_S$. The local temperature $T$ must at least temporarily
decrease below this threshold in order to provide the necessary
supersaturation for nucleation.

\begin{figure}[t]
\begin{center}
\epsfig{file=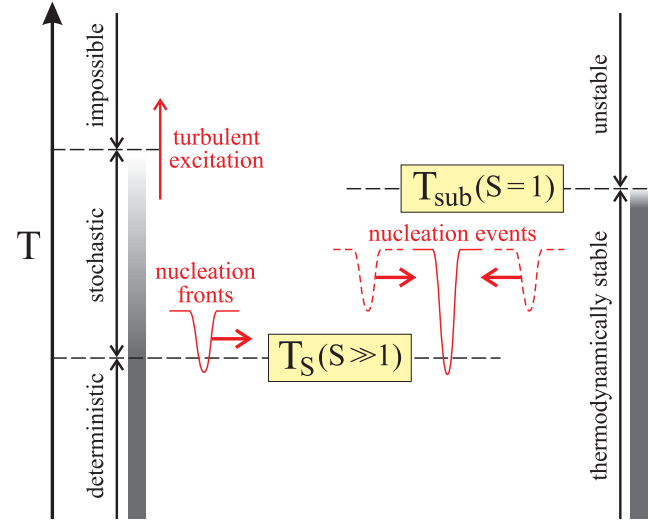, scale=0.65}
\caption[]{Regimes of turbulent dust formation. $T_S$: nucleation
  threshold temperature (supersaturation $S\ll 1$ required), 
  $T_{\rm sub}$: sublimation temperature.}
\label{fig:regimes}
\end{center}
\end{figure}

Depending on the relation between the local mean temperature
$\overline{T}$ and $T_S$, three different regimes can be distinguished
(see l.h.s. of Fig.~\ref{fig:regimes}): (i) the {\it deterministic}
regime ($\overline{T}<T_S$) where dust forms anyway, (ii) the {\it
stochastic} regime ($\overline{T}>T_S$) where $T<T_S$ can only be
achieved locally and temporarily by turbulent temperature
fluctuations, and (iii) a regime where dust formation is {\it
impossible}. The size of the stochastic regime depends on the
available turbulent energy. This picture of {\it turbulent dust
formation} is quite different from the usually applied {\it
thermodynamical picture} (r.h.s. of Fig.~\ref{fig:regimes}) where dust
is simply assumed to be present whenever $T<T_{\rm sub}$, where
$T_{\rm sub}$ is the sublimation temperature of a considered dust
material.

The investigations have been performed in the mesoscopic scale
regime, where the test volume is excited by a spectrum of waves within
a limited $k$-interval at given energy distribution, which are
generated at the boundaries (pseudo-spectral method for driven
turbulence). Two basic processes are found to be capable to
initiate the dust formation even in a host-hostile environment:
\begin{itemize}
\item[ 1)] expansion waves with $\overline{T}-\Delta T<T_S$ 
           ({\it nucleation fronts}).
\item[ 2)] interactions of two or more expansion waves which cannot
           produce sufficiently low temperatures for themselves, but
           the superposition of such waves can $\overline{T}-\Delta T_1
           - \Delta T_2 -\ldots<T_S$ ({\it nucleation events}).
\end{itemize}
After initiation, the dust condensation process is completed by a
phase of active particle growth until the condensible elements are
consumed, thereby preserving the dust particle number density for long
times.  However, radiative cooling (as follow-up effect) is found to
have an important influence on the subsequent dust formation, if 
the dust opacity reaches a certain critical value. This cooling
leads to a decrease of $\overline{T}(t)$ which may re-initiate the
nucleation. This results in a runaway process (unstable feedback loop)
until radiative and phase equilibrium is achieved. Depending on the
difference between the initial mean temperature $\overline{T}(t\!=\!0)$
and the radiative equilibrium temperature $T_{\rm RE}$, a considerable
local temperature decrease and density increase occurs. Since the turbulent
initiation of the dust formation process is time-dependent and
spatially inhomogeneous, considerable spatial variations of all
physical quantities (hydro-, thermodynamics, dust) occur during the
short time interval of active dust formation (typically a few seconds
after initiation), which actually {\it creates new turbulence}.

Thus, small turbulent perturbations have large effects in dust
forming systems. A convectively ascending, initially dust free gas
element, which is slightly warmer than its surroundings, can be
excited to form dust by waves running through it, even at otherwise
dust-hostile temperatures. The newly created dust particles may cause
the substellar atmosphere to become almost instantaneously optically
thick. Our 2D simulations show that the dust appears in lane-like and
curled structures. Small-scale dust structures merge and form larger
structures. Vortices appear to be present preferentially in regions
without or with only little dust. Non of these structures would
occur without turbulent excitation.

\begin{acknowledgements}
  The referee is thanked for the useful advises to the manuscript.
  Dipl.-Ing.~H.~Schmidt and Dr.~N.~Botta are thanked for discussion on
  the boundary problem and Dipl.-Ing.~M.~M\"unch for discussions on
  the Klein-HD-Code. This work has been supported by the \emph{DFG}
  (grants SE 420/19-1\&2, Kl 611/7-1, Kl 611/9-1).
\end{acknowledgements}

\begin{appendix}
\section{Analysis of characteristic numbers \& characteristic values 
         of test calculations}\label{append:analysischarnumb}

\begin{table*}[]
\begin{center}
\caption[]{Characteristic numbers and reference values used to
  analyse the chemistry and physics in the mesoscopic scale regime.
  Hydro- and thermodynamic reference values have been guided by static
  brown dwarf model atmosphere calculations
  (Tsuji~2002)\nocite{tsu2002b}. The reference values for the dust
  complex have been adopted from the experiences gained in Paper~I.
  Considered is the TiO$_2$ seed formation and growth by SiO and
  TiO$_2$.  ($\gamma = 7/5$, $a_{\rm l}^{\rm TiO_2}=1.95\cdot
  10^{-7}$cm, $m_{\rm H}$ mass of the hydrogen atom, $\sigma$
  Stefan-Boltzmann constant, $k$ Boltzmann constant, $\mu_{\rm kin}$
  kinematic viscosity [g cm$^{-1}$s$^{-1}$], $N_l=1000$)}
\label{tab:kennz}
\begin{small} \noindent\hspace*{-0.2cm}
\begin{tabular}{|c|ll|rcll|}
\hline
{\bf Name} &\multicolumn{2}{c|}{\bf Characteristic} 
                                        &\multicolumn{3}{c}{\bf Value} & \\
           &\multicolumn{2}{c|}{\bf Number} & inside & & outside & \\
\hline
Reynolds number & \multicolumn{2}{c|}{
          ${\bf \rm Re}=\frac{v_{\rm ref}l_{\rm ref}
          \rho_{\rm ref}}{\mu_{\rm kin}}$}  & & & &\\
                & \multicolumn{2}{c|}{
          \hspace{2.5cm}$=5.88\cdot10^{-6} \sqrt{T}\qquad^{(*)}$} & 
          $1.2\,10^{10}$ & $\ldots$ & $1.2\,10^{7}$ & \\

Mach number & \multicolumn{2}{c|}{
          ${\bf\rm M }=\frac{v_{\rm ref}}{c_{\rm s}}$}
          & & 0.1 & & \\

Froude number & \multicolumn{2}{c|}{
          ${\bf\rm Fr}=\frac{v_{\rm ref}^2}{l_{\rm ref}}
          \frac{1}{g_{\rm ref}}$}
          & 0.11 & & 0.05 &\\
\hline
Radiation number & \multicolumn{2}{c|}{
         ${\bf\rm Rd}= 
          4\kappa_{\rm ref}\sigma T^4_{\rm ref} \cdot 
          \frac{t_{\rm ref}}{ P_{\rm ref}}$} & 
           $0.207$& $\ldots$ & $8.6\, 10^{-3}$ & $^{(\Diamond)}$\\
              & & & $ 2.07 $ & $\ldots$ & $0.86$ &
                $^{(\Diamond\Diamond)}$\\
\hline
Damk\"ohler number of nucleation & \multicolumn{2}{c|}{
         ${\bf\rm Da^{\rm nuc}_{\rm d}} =
          \frac {t_{\rm ref} J_{*, {\rm ref}} }
          {\rho_{\rm ref}L_{0, {\rm ref}}}$} & 
                   0 & $\ldots$ & $3.24\,10^5$ & \\[0.8ex]

Damk\"ohler number of growth & \multicolumn{2}{c|}{
         ${\bf\rm Da^{\rm gr}_{\rm d}} =
                    \frac{t_{\rm ref}\chi_{\rm ref}}
                   {(\frac{4\,\pi}{3}\langle a \rangle^3_{\rm ref})^{1/3}}$} & 
                   $9.78\,10^4$ & $\ldots$ &  $97.8$ & \\

                      & & & & \multicolumn{2}{l} {$j=0$ : $1$} & \\[-1.5ex]
Sedlma\"yr number $(j \in \mathbb{N})$ & \multicolumn{2}{c|}{
                   ${\bf\rm Se_{\rm j}} =
                   \left( \frac{a_l}{\langle a \rangle_{\rm ref}}\right)^{\rm j}$}
                          & & \multicolumn{2}{l} {$j=1$ : $0.195$} & \\[-1ex]
                      & & & & \multicolumn{2}{l} {$j=2$ : $0.0381$}& \\[-0.2ex]
                      & & & & \multicolumn{2}{l} {$j=3$ : $7.44\,10^{-3}$}& \\
\hline
Element Consumption number & \multicolumn{2}{c|}{
                   ${\bf\rm El} = 
                   \frac{\rho_{\rm ref} L_{0, {\rm ref}}N_l}
                   {n_{\rm <H>, ref} \epsilon_{\rm ref}}$} & &$7.27\,10^{-4}$ & & \\
\hline
\hline
{\bf Name}& \multicolumn{2}{c|}{\bf Reference Value} 
          & \multicolumn{3}{c}{\bf Value} &  \\
& & & inside & & outside & \\
\hline
temperature       & $T_{\rm ref}$ & [K] & 
                     2200 & $\ldots$ & 1000 & \\
density           & $\rho_{\rm ref}$ & [g/cm$^{3}$] & 
                     $10^{-3}$ & $\ldots$ & $10^{-6}$ & \\
thermal pressure  & $P_{\rm ref} = \frac{kT_{\rm ref}\rho_{\rm ref}}
                    {2.3m_{\rm H}}$ & [dyn/cm$^2$] &  
                     $7.78\,10^7$ & $\ldots$ & $3.54\,10^4$  & \\
velocity of sound & $c_{\rm S}=\sqrt{\gamma\frac{P_{\rm ref}}
                    {\rho_{\rm ref}}}$  & [cm/s] & 
                     $3.3\,10^5$ & $\ldots$ & $2.23\,10^5$ & \\
velocity          & $v_{\rm ref}$ & [cm/s] & 
                     & $\approx c_{\rm S}/10$ & & \\
length            & $l_{\rm ref}$ & [cm] & 
                     & $10^5$ & &  \\
hydrodyn. time    & $t_{\rm ref} = \frac{l_{\rm ref}}{v_{\rm ref}}$ & [s] & 
                     $3.03$ & $\ldots$ &4.49& \\
gravitational acceleration  
                  & $g_{\rm ref}$ & [cm/s$^2$] & 
                     & $10^5$ & & \\
\hline
total absorption coefficient 
                  & $\kappa_{\rm ref}$ & [1/cm] & 
                     $10^{-3}$ & $\ldots$ & $3\cdot 10^{-7}$ 
                     & $^{(\Diamond)}$  \\
              & & & $0.1$ & $\ldots$ & $3\cdot 10^{-5}$ 
                    &  $^{(\Diamond\Diamond)}$ \\
\hline
nucleation rate   & $J_{*, \rm ref}/n_{\rm <H>, ref}$ & [1/s] & 
                     $0$ & $\ldots$ & $2.5\cdot 10^{-6}$ &  \\
0$^{\rm th}$~dust moment ($=n_{\rm d}/\rho$)
                  & $L_{0, \rm ref}$ & [1/g] & 
                  & $\ldots$ & $1.35\,10^{14}$ 
                     & $^{(\Box)}$ \\
heterog. growth velocity 
                  & $\chi_{\rm ref}$ & [cm/s] & 
                     $\lesssim 0$ & $\ldots$ & $3.51\,10^{-5}$ &  \\
mean particle radius 
                  & $\langle a \rangle_{\rm ref}$ & [cm] & 
                     & $10^{-6}$ & & \\
\hline
element abundance & $\epsilon_{\rm ref}$ & [--] & 
                     & $10^{-6}$ & & \\
total hydrogen density
                  & $n_{\rm <H>, ref} = \frac{\rho_{\rm ref}}
                    {1.4\,m_{\rm H}}$ & [1/cm$^3$] & 
                     $4.22\, 10^{20}$  & $\ldots$ & $4.22\,10^{17}$ & \\
SiO density &  $n_{\rm <SiO>, ref}$ & [1/cm$^3$] & $1.50\,10^{16}$ & $\ldots$ & $1.50\,10^{13}$&\\ 
TiO$_2$ density &  $n_{\rm <TiO_2>, ref}$ & [1/cm$^3$] & $4.12\,10^{13} $ & $\ldots$ & $4.10\,10^{10}$&\\ 
\hline
\end{tabular}
\end{small}\\
$^{(*)}$ {\footnotesize For a H$_2$/He rich gas (Eq.~(10) in
Woitke\plus Helling~2003a\nocite{wh2003a}).}\quad $^{(\Diamond)}$ {\footnotesize gas only}
$^{(\Diamond\Diamond)}$ {\footnotesize dust and gas } $^{(\Box)}$
{\footnotesize Gail \& Sedlmayr~(1999)\nocite{gs99}}
\end{center}
\end{table*}

\begin{table*}[]
\begin{center}
\caption[]{Characteristic numbers and reference values for the test calculations carried out in the previous sections. For definition of characteristic numbers see Table~\ref{tab:kennz}}
\label{tab:kennzalle}
\begin{small} \noindent\hspace*{-0.2cm}
\begin{tabular}{|cl|l|l|l|l|}
\hline
 & & {\bf A} & {\bf B} & {\bf C} & {\bf A$^{\prime}$}\\
\hline
\multicolumn{6}{|c|}{\underline{Hydro- \& Thermodynamic}}\\
$T_{\rm ref}$    & [K]             & 2100 & 2100 & 1900 & 2100\\
$\rho_{\rm ref}$ & [g\,cm$^{-3}$]  
& $3.16\,10^{-4}$&  $3.16\,10^{-4}$& $8.13\,10^{-4}$ & $3.16\,10^{-4}$\\
$l_{\rm ref}$    & [cm]            
& $10^5$ & $10^5$ & $10^5$ & $10^5$\\
$v_{\rm ref}$    & [cm\,s$^{-1}$]  
& $3.25\,10^4$ & $6.50\,10^3$ & $3.09\,10^4$ & $3.25\,10^5$\\
$g_{\rm ref}$    & [cm\,s$^{-2}$]  & $10^5$ & $10^5$ & $10^5$ & $10^5$\\
$T_{\rm RE}$     & [K]             
& $1980$ $(=0.9T_{\rm ref})$ &$1980$ $(=0.9T_{\rm ref})$ & $1710$ $(=0.9T_{\rm ref})$ & $1980$ $(=0.9T_{\rm ref})$ \\
$\kappa_{\rm ref}$  & [1/cm]       & $2\,10^{-5}$ & $2\,10^{-5}$ & $2\,10^{-5}$  & $2\,10^{-5}$\\
\hline
\multicolumn{6}{|c|}{\underline{Dust \& Chemistry}}\\
$J_{\rm *, ref}$ & [s$^{-1}$cm$^{-3}$] & $10^{12}$ & $10^{12}$ & $10^{12}$ & $10^{12}$  \\
$a_{\rm ref}$    & [cm]                & $10^{-6}$ & $10^{-6}$ & $10^{-6}$ & $10^{-6}$\\
$\chi_{\rm ref}$    & [cm\,s$^{-1}$]   & $10^{-2}$ & $10^{-2}$ & $10^{-2}$ & $10^{-2}$\\
$L_{\rm 0, ref}$ & [g$^{-1}$]          & $10^{10}$ & $10^{10}$ & $10^{10}$ & $10^{10}$\\ 
\hline
\multicolumn{6}{|c|}{\underline{Derived reference values}}\\
p$_{\rm ref}$    & [dyn\,cm$^{-2}$]  & $2.38\,10^7$ & $2.38\,10^7$ & $5.54\,10^7$ & $2.38\,10^7$\\
t$_{\rm ref}$    & [s]               & 3.08 & 15.4 & 3.24 & 0.3\\
$L_{\rm j, ref}$ & [cm$^{\rm j}$g$^{-1}$]& $(\frac{4\pi\,a_{\rm ref}^3}{3})^{j/3}L_{\rm 0, ref}$ 
& $(\frac{4\pi\,a_{\rm ref}^3}{3})^{j/3}L_{\rm 0, ref}$  
& $(\frac{4\pi\,a_{\rm ref}^3}{3})^{j/3}L_{\rm 0, ref}$  
& $(\frac{4\pi\,a_{\rm ref}^3}{3})^{j/3}L_{\rm 0, ref}$\\[0.1cm]  
\hline
\hline
\multicolumn{6}{|c|}{\underline{Hydro- \& Thermodynamic}} \\
$M$  & &  1/10  &  1/50  & 1/10 & 1\\ 
$Fr$ & & 0.105  &  $4.22\,10^{-3}$ & 0.095 & 10.5\\
$Rd$ & & $1.139\,10^{-2}$ & $5.70\,10^{-2}$ & $3.45\,10^{-3}$ & $1.14\,10^{-3}$\\
\hline
\multicolumn{6}{|c|}{\underline{Dust \& Chemistry}}\\
$Da^{\rm nuc}_{\rm d}$&& $9.73\,10^5$ & $4.87\,10^6$ & $3.98\,10^5$ & $9.73\,10^4$\\
$Da^{\rm gr}_{\rm d}$&& $1.91\,10^4$  & $9.55\,10^4$ & $2.01\,10^4$ & $1.91\,10^3$\\
 $Se_0$ &&  1.0   & 1.0   & 1.0  & 1.0\\
 $Se_1$ &&  0.195 & 0.195 &0.195 & 0.195 \\
 $Se_2$ &&  $3.8\,10^{-2}$ &$3.8\,10^{-2}$ &$3.8\,10^{-2}$ & $3.8\,10^{-2}$\\
 $Se_3$ &&  $7.41\,10^{-3}$ & $7.41\,10^{-3}$ & $7.41\,10^{-3}$ & $7.41\,10^{-3}$\\
 $El$   &&  $1.41\,10^{-7}$ & $7.07\,10^7$ & $3.82\,10^{-7}$ & $1.41\,10^{-7}$\\
\hline
\hline
\multicolumn{6}{|c|}{\underline{Turbulence}}\\
$\varepsilon$  (Eq.~\ref{equ:energydiss}) & [cm$^2$ s$^{-3}$]& $2.40\,10^8$ & $1.92\,10^6$ & $2.06\,10^8$ & $2.41\,10^{11}$\\
$S_{\rm ref}$  (Eq.~\ref{equ:Temp}) & [erg/K] 
& $1.42\,10^9$ & $1.37\,10^9$ &  $1.28\,10^9$ & $1.4\,10^9$\\
$k_{\rm min}=2\pi/l_{\rm max}$  & [1/cm] 
& $6.30\,10^{-5}$ & $6.30\,10^{-5}$ & $6.30\,10^{-5}$ & $6.30\,10^{-5}$\\
$k_{\rm max}=2\pi/(3\,dx)$      & [1/cm] 
& $2.10\,10^{-2}$ & $2.10\,10^{-2}$ & $2.10\,10^{-2}$ & $5.30\,10^{-3}$\\
$l_{\rm max}$                   & [cm]   & $l_{\rm ref}$ & $l_{\rm ref}$ & $l_{\rm ref}$ & $l_{\rm ref}$  \\
$dx$  for $N_x=500$  (1D)     &  [cm]   & 
$1.00\,10^2$  & $1.00\,10^2$ & $1.00\,10^2$ & $-$\\ 
$dx$  for $N_x=128$  (2D)    &  [cm]   & 
$-$  & $-$ & $-$ & $3.94\,10^2$\\ 
\hline
\end{tabular}
\end{small}

\end{center}
\end{table*}

\end{appendix}


\newcommand{\etalchar}[1]{$^{#1}$}

\end{document}